\documentclass[twocolumn,graphics,floatfix,a4paper,prb,nofootinbib]{revtex4}
\usepackage{graphicx, epsfig,amsmath, amssymb,mathtools,braket,bbm,turnstile,color,fancyhdr,marvosym,bm}
\usepackage{tikz}
\usepackage[strict]{changepage}
\usepackage{calc}
\newcommand{\Fig}[1]{Fig.~\ref{#1}}

\newcommand{\martin}[1]{\textcolor{red}{}}
\newcommand{\Eq}[1]{Eq.(\ref{#1})}

\newcommand{\bit}{\begin{itemize} \setlength{\itemsep}{0ex} \setlength{\topsep}{0ex} }
\newcommand{\eit}{\end{itemize}}
\newcommand{\be}{\begin{equation}}
\newcommand{\ee}{\end{equation}}
\newcommand{\bea}{\begin{eqnarray}}
\newcommand{\eea}{\end{eqnarray}}
\newcommand{\ba}{\begin{align}}
\newcommand{\ea}{\end{align}}
\newcommand{\id}{\mathbbm{1}}

\newcommand{\SKIP}[1]{}

\def \ra{\rightarrow}

\def \la{\leftarrow}

\def \eps{\epsilon}

\def \tr{{\rm tr}}

\begin{document}
\title{Continuous Matrix Product States for Inhomogeneous Quantum Field Theories:
  a Basis-Spline Approach}
\author{Martin Ganahl}
\affiliation{Perimeter Institute for Theoretical Physics, 31 Caroline Street North, Waterloo, ON N2L 2Y5, Canada}
\email{martin.ganahl@gmail.com}
\begin{abstract}
Continuous Matrix Product States (cMPS) are powerful variational ansatz states for ground states 
of continuous quantum field theories in (1+1) dimension. In this paper we introduce a novel 
parametrization of the cMPS wave function based on basis-spline functions, which we coin spline-based MPS (spMPS), 
and develop novel regauging techniques for inhomogeneous cMPS.
We extend a recently developed ground-state optimization algorithm for 
translational invariant cMPS [M. Ganahl, J. Rinc\'on, G.Vidal. Phys.Rev.Lett. 118,220402 (2017)]
to the case of inhomogeneous cMPS and, as proof-of-principle,
use it to obtain the ground-state of a gas of Lieb-Liniger
bosons in a periodic potential.
The proposed method provides a first working implementation of a cMPS optimization for 
non-translational invariant continuous Hamiltonians.
\end{abstract}

\maketitle

\section{Introduction}

One of the big challenges in theoretical physics is the treatment of interacting
quantum theories of many particles. Strong interaction between particles, such as
electron-electron repulsion, electron-phonon interactions, or magnetic exchange interactions 
between spin degrees of freedom, are considered to be fundamental to phenomena like e.g. 
high-$T_C$ superconductivity \cite{bednorz_possible_1986} or the fractional quantum Hall effect 
\cite{tsui_two-dimensional_1982,laughlin_anomalous_1983}.
In most cases, however, obtaining exact solutions of these models is far beyond current
capabilities, and a big effort has been devoted to develop approximate analytic and numerical
tools to tackle these system. For 1-dimensional (1d) quantum systems, 
so called Matrix Product States \cite{fannes_finitely_1992,white_density_1992}, 
have proven to be particularly powerful tools to obtain ground-states \cite{white_density_1992} 
and also low-lying excited states
\cite{pirvu_matrix_2012,haegeman_variational_2012,vanderstraeten_scattering_2015}
of quantum lattice systems.
In particular, the Density Matrix Renormalization Group 
\cite{white_density_1992,white_density-matrix_1993} (DMRG),
a ground state optimization method within the variational class of MPS,
is today considered the most powerful method for obtaining ground-states of 1d quantum lattice
system. Over the last few years, growing effort has been devoted to extend
the DMRG to the case of continuous quantum field theories
\cite{stoudenmire_one-dimensional_2012,baker_one-dimensional_2015,baker_erratum:_2016,
  stoudenmire_sliced_2017,baker_chemical_2017} using lattice discretization techniques.

Recently, tensor network methods have been extended from lattice systems to continuous 
quantum field theories \cite{verstraete_continuous_2010,haegeman_entanglement_2013}, among which
so called continuous Matrix Product States (cMPS) have received growing attention 
during the last few years 
\cite{haegeman_applying_2010,draxler_particles_2013,quijandria_continuous_2014,chung_matrix_2014,
  quijandria_continuous-matrix-product-state_2015,haegeman_quantum_2015,
  rincon_lieb-liniger_2015,chung_matrix_2015,
  draxler_atomtronics_2016,ganahl_continuous_2017}. 
Such methods have interesting applications ranging from cold atomic gases \cite{cazalilla_one_2011,bloch_ultracold_2005} 
to quantum chemistry to applications even in holography and quantum gravity 
\cite{swingle_entanglement_2012,miyaji_continuous_2015}. 
For translational invariant theories, efficient methods 
have been developed to obtain accurate ground-state wave functions for infinite systems
\cite{ganahl_continuous_2017,haegeman_time-dependent_2011} or systems on a ring 
\cite{draxler_atomtronics_2016}, as well as low-lying excited states 
\cite{draxler_particles_2013}. All these methods work directly in the continuum, without
the need of introducing any discretization.
For the case of inhomogeneous Hamiltonians, first important formal and
perturbative results have been obtained \cite{haegeman_quantum_2015}. However,
despite a considerable ongoing effort, no successful optimization
for inhomogeneous cMPS has been achieved so far.
The present manuscript aims at filling this gap.
The main contribution of this paper is the 
introduction of a spline-based representation for inhomogeneous cMPS 
which we coin spline-based Matrix Product States (spMPS). 
In the following, we will discuss in depth the computational advantages of 
this representation, and will develop new tools for manipulating and optimizing cMPS. 
In particular, we will present regauging algorithms for 
cMPS in spMPS form, and an extension of a recently proposed optimization 
algorithm for translational invariant Hamiltonians \cite{ganahl_continuous_2017}
to the case of periodic spMPS.
We will conclude by presenting results for a spMPS
ground state of a gas of interacting Lieb-Liniger bosons in a periodic potential, and will
discuss possible future applications.

\section{Continuous Matrix Product States and basis-spline interpolation}
The setting we choose in this manuscript is a gas of a single species
of bosons on the real line with a periodic unit-cell of length $L$, which we
choose to be ${L=}1$.
For such a system, a generic cMPS wave function assumes the form
\begin{align}
  \ket{\Psi}=v_l\mathcal{P}e^{\int_{-\infty}^{\infty}dx\; Q(x)\otimes\mathbbm{1}+R(x)\otimes\psi^{\dagger}(x)}v_r\ket{0},
\end{align}
where $Q(x),{R(x)\in} \mathbbm{C}^{D\times D}$ are periodic matrix functions with period $L$, 
and $\psi^{\dagger}(x)$ is a bosonic creation operator
for the vacuum state $\ket{0}$. $v_l$ and $v_r$ are arbitrary boundary vectors at $x=\pm\infty$,
and $\mathcal{P}$ is the path-ordering operator.
For given $Q(x), R(x)$, any local observable $\braket{\Psi|O(x)|\Psi}$, e.g. $O(x)=\frac{d\psi^{\dagger}(x)}{dx}\frac{d\psi(x)}{dx}$, 
can be calculated using 
contraction techniques similar to the lattice case. In particular, the expectation value of 
energy {\it densities} can be evaluated. 
This enables the application of the variational principle to the class of cMPS wave functions, 
and for homogeneous systems has lead to the development of 
efficient methods for calculating ground-state wave functions of continuous quantum field 
theories \cite{ganahl_continuous_2017,haegeman_time-dependent_2011}.

In numerical applications, the functions $Q(x)$ and $R(x)$ are usually not known analytically.
Instead of using continuous functions $Q(x), R(x)$, a common approach in numerical 
mathematics is to define a set of points $(x_i,Q(x_i))$ and $(x_i,R(x_i))$ on a 
fine grid $x_i, {i=}1\dots M_{grid}$, and use these points to 
approximate continuous functions $Q(x),R(x)$. 
When aiming to approximate the ground state of a continuous Hamiltonian $H$, one then
uses the same discretization for the Hamiltonian.
The number $M_{grid}$ of grid points should be chosen according to the smallest
scale of variation in any of the Hamiltonian parameters. For example, if the
Hamiltonian contains a chemical potential of cosine form with a period of one,
$\mu(x)\sim\cos(2\pi x)$, one might
expect the same period (and possibly higher harmonics) 
to be present in the ground state wave function.
Thus, $M_{grid}$ has to be chosen so large that at least this variation
may be well resolved. This discretization, however, introduces an error in the 
evaluation of expectation values (see discussion below).
In this paper, we propose a new parametrization of the cMPS wave function
and combine it with a recently proposed optimization method for
homogeneous cMPS. Our central proposal is to use basis-spline ({\it b-spline} 
in the following) interpolation to parametrize the continuous
functions $Q(x),R(x)$.
For a detailed introduction to b-splines we refer the reader to the excellent review 
\cite{bachau_applications_2001}. 
In the following, we will give a short introduction to b-splines and
summarize their most important properties.

\subsection{An introduction to basis-splines}
\begin{figure}
  \includegraphics[width=1.0\columnwidth]{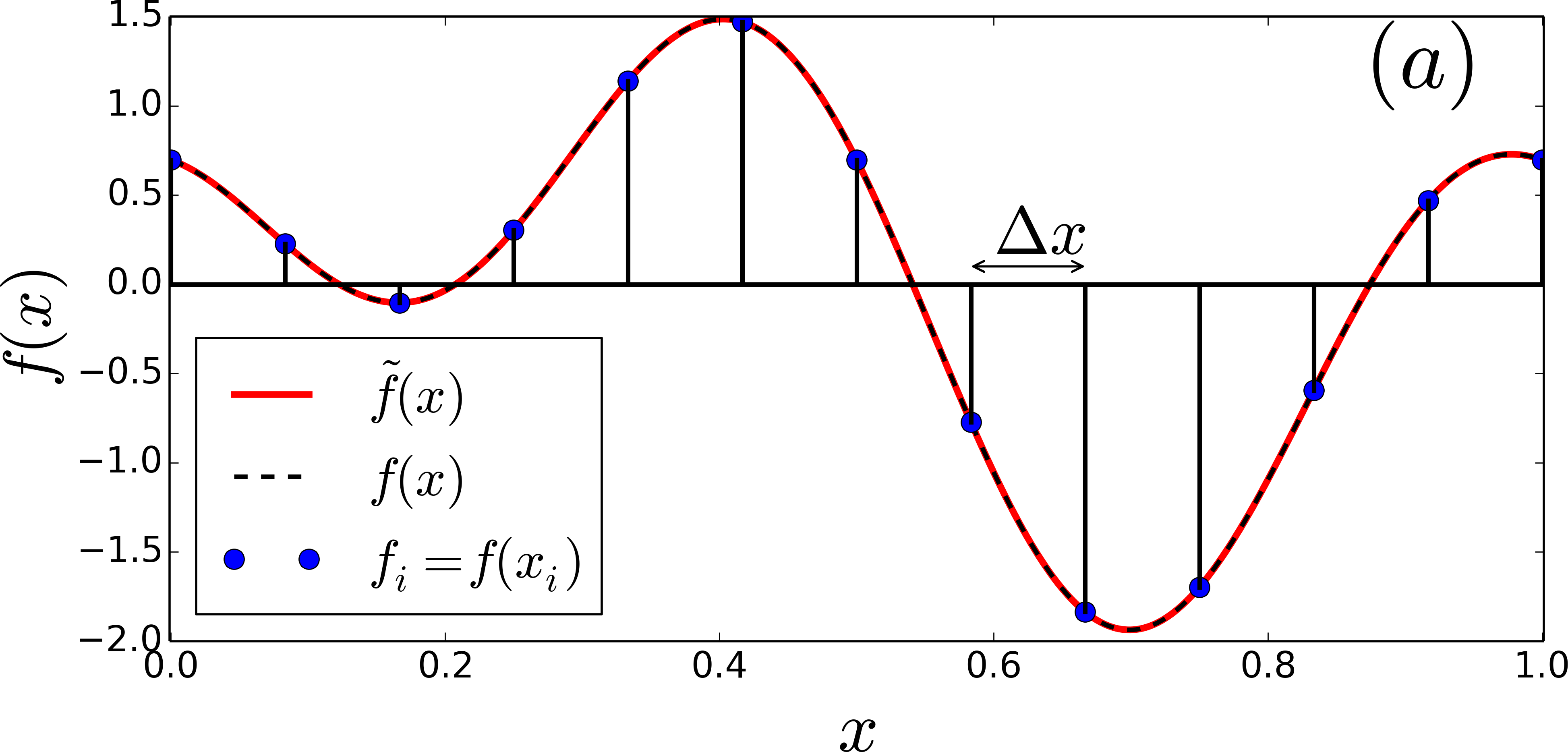}
  \includegraphics[width=0.48\columnwidth]{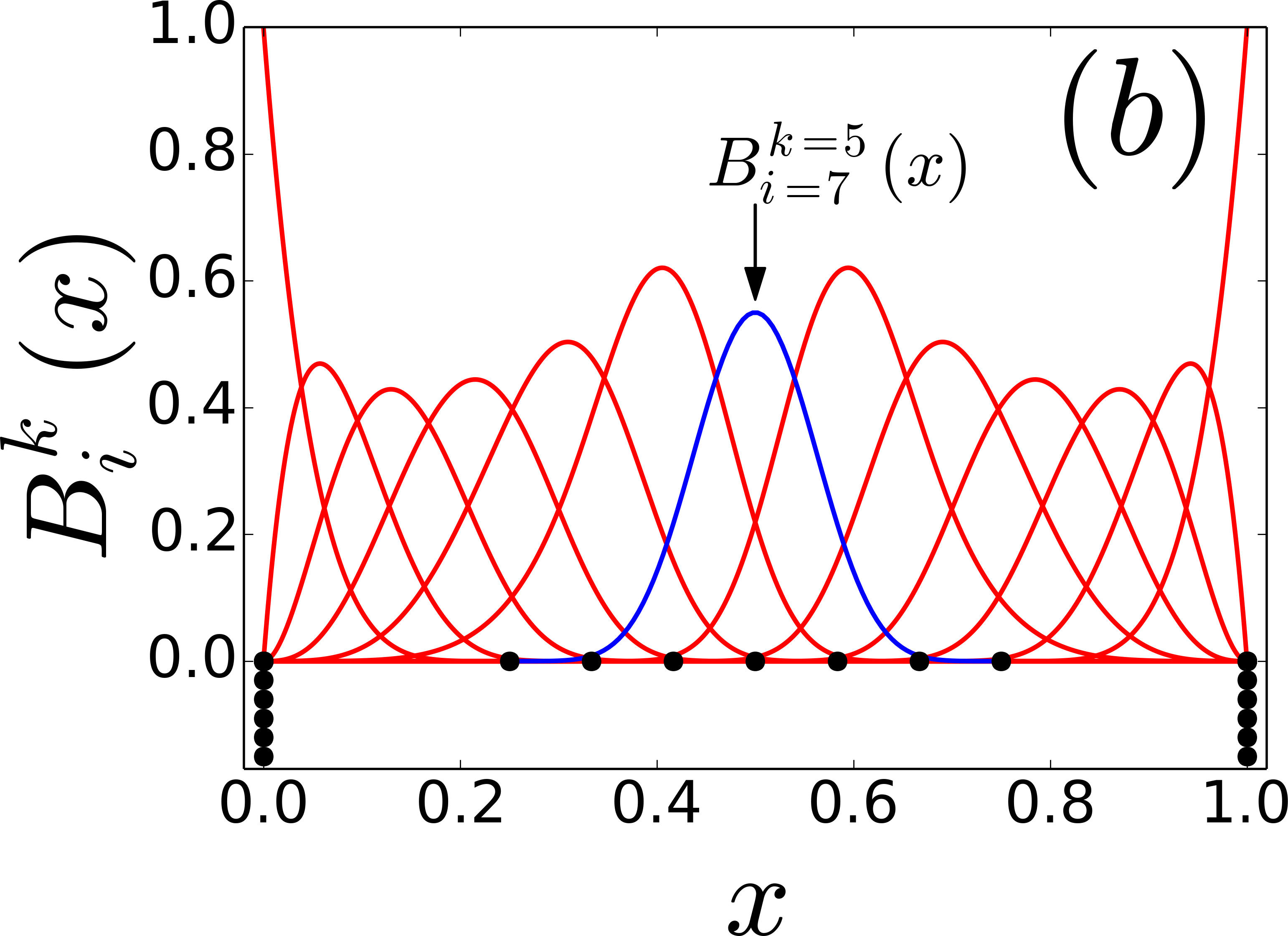}
  \includegraphics[width=0.48\columnwidth]{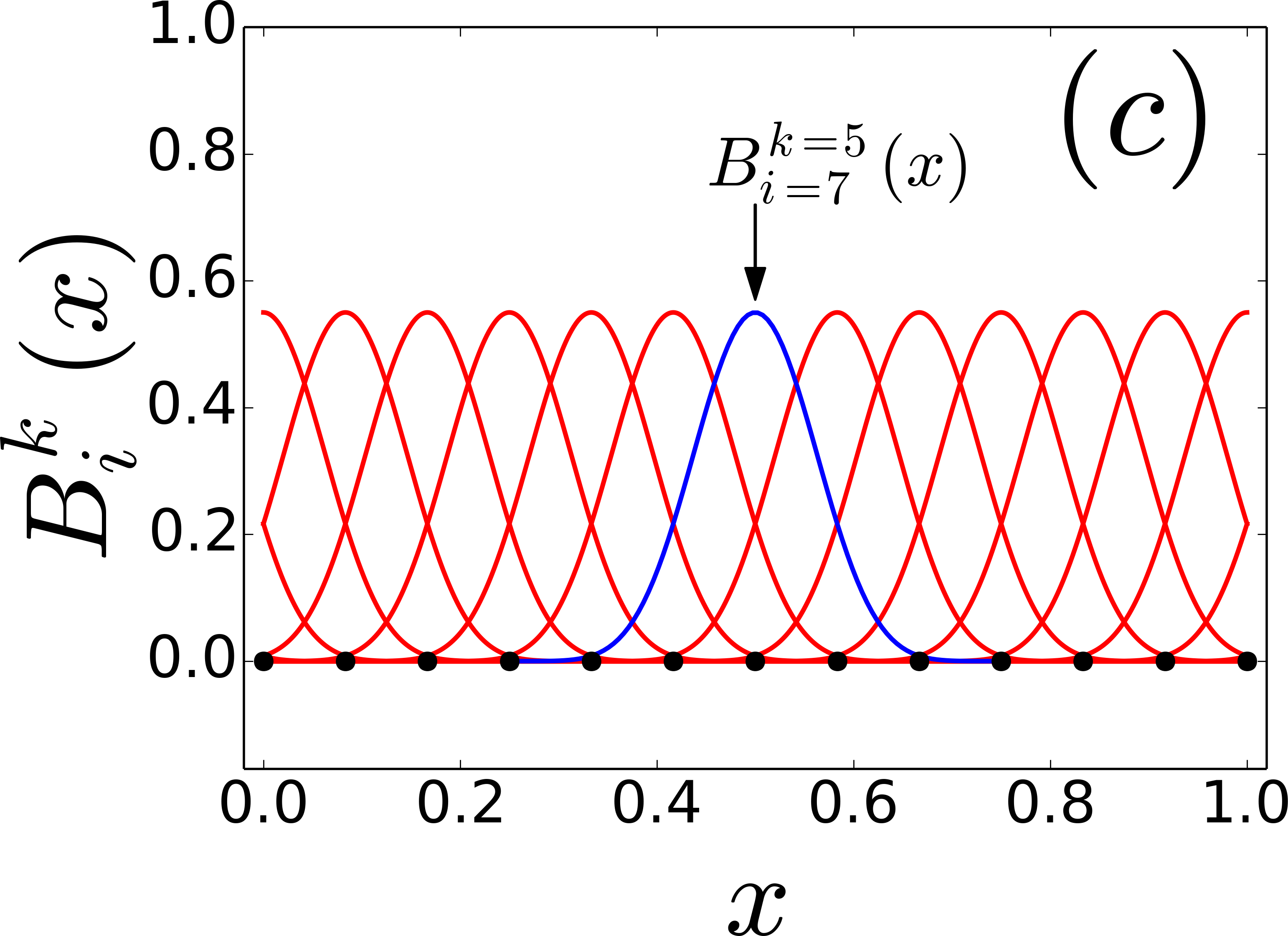}
  \caption{{\bf Example of b-spline interpolation} (a) We generate data-points $(x_i,f_i=f(x_i))$
    (blue dots) using the values $f(x_i)$ of \Eq{eq:testfun} at 13 equally 
    spaced points $x_i\in[0,1]$ with spacing $\Delta x$. 
    We then use a b-spline interpolation to 
    obtain the function $\tilde f(x)$ (red solid line). For comparison we also 
    show $f(x)$ (black dashed line). (b) B-spline polynomials 
    $B^k_i(x)$ for spline interpolation with open boundary conditions. 
    (c) B-spline polynomials $B^k_i(x)$ for spline interpolation with 
    periodic boundary condition. We have highlighted the polynomial $B^5_7(x)$ in blue.}\label{fig:bspline}
\end{figure}

Given a set of $P$ discrete data points $(x_i,f_i)$
with $x_i,f_i\in \mathbbm{R}$, and $x_1<x_2\dots<x_P$, a frequently encountered task 
in numerical analysis is to 
find a smooth curve $\tilde f(x)$ which runs through all points $(x_i,f_i)$, 
i.e $\tilde f(x_i)=f_i$. 
Smooth means that the curve should be differentiable
up to some order $p$ everywhere inside the domain $x_1<x<x_P$, with the possible
exception of the boundary points $x_1,x_P$. The simple yet powerful idea behind 
b-spline interpolation is to write $\tilde f(x)$ as a piece-wise polynomial function.
Piece-wise polynomial here means that
$\tilde f(x)$ has an expansion in a set of polynomials $B_i^k(x)$ of degree $k$,
\begin{align}
  \tilde f(x)=\sum_{i=1} \beta^iB_i^k(x), x\in [x_1,x_P]\label{eq:bspline},
\end{align}
where every $B_i^k(x)$ is non-zero only inside a small region, and zero everywhere else,
and $\beta^i$ are expansion coefficients.
The $B_i^k(x)$ are designed to be sufficiently often differentiable, such 
that $\tilde f(x)$ has the desired smoothness properties. The coefficients
$\beta^i$ are chosen such that $\tilde f(x_i)=f_i$.
\Fig{fig:bspline} (a) shows an example of a simple b-spline interpolation. We 
chose thirteen equally spaced points $x_i\in [0,1]$ and generated
the data points $(x_i,f_i=f(x_i))$ from the function 
\begin{align}
  f(x)=\sin(2\pi x)+\cos(4\pi x+0.8)\label{eq:testfun}.
\end{align}
We then used a degree $k=5$ interpolation to obtain the function $\tilde f(x)$.
All three are plotted in \Fig{fig:bspline}(a). In particular, the function $\tilde f(x)$
is seen to approximate $f(x)$ very well, with a relative error less than $0.1\%$.
In \Fig{fig:bspline}(b) we plot the corresponding polynomials $B^k_i(x)$, and,
additionally, the so called {\it knot-points} $t_j$ \cite{bachau_applications_2001}
as black dots. The $\{t_j\}$ are a set of ascending, not necessarily distinct
points with $x_1\leq t_j\leq x_P$.
If a point $t_j$ appears multiple ($\nu_j$) times, it is said to have multiplicity $\nu_j$.
The knot points $t_j$ and their multiplicities $\nu_j$
are part of the definition of the $B_i^k(x)$.
The $t_j$ are determined from the points $x_i$, and for standard applications, 
their multiplicity is chosen to be $\nu_1=\nu_P=k+1$ and $\nu_j=1$ for $j=2\dots P-1$. We have 
indicated the multiplicity of the points $t_j$ in \Fig{fig:bspline}(b) accordingly.
As can be seen from \Fig{fig:bspline}(b), a polynomial $B_i^k(x)$  
has support on the interval $]t_i,t_{i+k+1}[$, and vanishes outside of it. This
is illustrated for $B^5_7(x)$, highlighted in blue. Note that for $0<x<1$,
all polynomials, together with their derivatives $\partial^n B_i^k(x),n=0,\dots,k-1$
are continuous (denoted by $B_i^k\in C^{k-1}$). At $x=0$, however, the first $k+1$ polynomials
are $B_i^k(x)\in C^{i-1}$, for ${i=}1,\dots {k+1}$ (and likewise for $x=1$). For example, 
$B^5_0(x)$ is discontinuous at $x=0$, $B_1^5(x)$ is continuous, but has
a discontinuous first derivative, a.s.o (and likewise $B^5_{13}(x),B^5_{12}(x),\dots$ at $x=1$).

So far we did not make use of the fact that $f(x)$ in \Eq{eq:testfun} is periodic. 
The b-spline polynomials $B^k_i(x)$ for a periodic expansion are shown in \Fig{fig:bspline}(c).
The knot points are in this case equidistantly spaced and 
chosen to have multiplicity one, and all $B_i^k(x)$ differ
only by a shift in $x$ and are $B_i^k(x)\in C^{k-1}$.
Using periodic b-spline interpolation has the advantage of giving a higher accuracy
when approximating a periodic function $f(x)$ 
with $\tilde f(x)$. In this paper, we will thus use periodic b-spline
interpolation throughout.
Most packages for numerical computation provide efficient routines to generate and 
handle b-spline interpolations. For this work, we use the routines
{\it splrep} and {\it splev} provided by the {\it scientific python} package.

\subsection{Spline-based Matrix Product States}
Returning to our previous discussion of cMPS, we propose to parametrize the continuous 
matrix functions $Q(x),R(x)$ as 
\begin{align}
  Q_{\alpha\beta}(x)&=\sum_i \mathcal{Q}^i_{\alpha\beta}B_i^k(x)\\
  R_{\alpha\beta}(x)&=\sum_i \mathcal{R}^i_{\alpha\beta}B_i^k(x), x\in [0,L].
\end{align}
The expansion coefficients $\mathcal{Q}^i_{\alpha\beta},\mathcal{R}^i_{\alpha\beta}$ contain
the variational parameters of our ansatz state, and
$\{B^k_i(x)\}$ are a given set of b-spline polynomials on an interval $I=[0,L]$.
Such a parametrization has several appealing features. For example, as we have illustrated above,
any sufficiently smooth function can be very accurately approximated by a small number
of basis spline functions. If the matrices $Q(x)$ and $R(x)$ are expected
to be sufficiently smooth, this parametrization is more efficient
than storing matrices $Q(x), R(x)$ on a very fine grid with $M_{grid}$ grid points.
Another appealing property is that the matrices $Q(x),R(x)$ can be evaluated at any point $x\in I$.
This is a significant advantage for example when considering the action of local operators on 
the cMPS $\ket{\Psi}$, as required in e.g. minimization algorithms. For example,
in the cMPS framework, the action of the operator
$\frac{d\psi(x)}{dx}\ket{\Psi}$ is translated into the operation $\frac{d\psi(x)}{dx}\ket{\Psi}\sim [Q(x),R(x)]+\frac{dR}{dx}$
on the matrices $Q(x),R(x)$ \cite{haegeman_calculus_2013}.
The derivative $\frac{dR}{dx}$ can be obtained {\it analytically} from the spline representation 
of the matrices. This has to be compared to the finite-difference approximation $\frac{dR}{dx}\approx\frac{R(x_{i})-R(x_{i-1})}{x_i-x_{i-1}}$
for a discrete set of points $(x_i,R(x_i))$.
As a potential benefit of the b-spline parametrization, 
discretization errors may be severely reduced, as compared to a finite difference approach for cMPS, 
or standard discretization methods using lattice MPS.

One of the main differences of our proposed approach 
as compared to regular lattice-MPS discretization methods is that in the latter case
one introduces a discretization of the Hamiltonian with 
a finite lattice spacing $a$, and an identical discretization for the wave function 
$\ket{\Psi}$. For the simplest discretization scheme, the error of observables
is of order $a$ throughout the calculation (higher order discretization of
derivatives usually yield more accurate results, at the cost of introducing longer ranged
terms in the discretized Hamiltonian).
In our proposed approach on the other hand,
the error is determined by how well the spline-interpolation can capture
the typical oscillations present in a periodic cMPS. For ground states, it is reasonable to expect
that these oscillations are of the same order as those present in the Hamiltonian parameters.
For example, if these oscillations are of the same order as those shown in \Fig{fig:bspline}(a), a
small number of b-spline polynomials will already suffice to give very accurate results.
In particular, the accuracy will be much higher 
than the naive lattice spacing $\Delta x$ used in the figure. This is evident
from the high accuracy with which the function $f(x)$ from \Eq{eq:testfun} is reproduced
by the b-spline interpolation $\tilde f(x)$ \Eq{eq:bspline}.

\section{Regauging an inhomogeneous cMPS in the thermodynamic limit}\label{sec:gauging}

Before proceeding, it is useful to introduce some diagrammatic notation at this point
and elaborate on its relation to regular lattice MPS diagrams.
For a gas of a single species of bosons (which we will be dealing with in this paper),
the cMPS can be thought of as a collection of matrices $A^{\sigma}(x)$ of the form
\begin{align}
  A^0(x)=\mathbbm{1+}\eps Q(x), \sigma=0\\
  A^{\sigma}(x)=\sqrt{\frac{\eps^{\sigma}}{\sigma!}} [R(x)]^{\sigma}, \sigma>0\nonumber
\end{align}
at any point $x$ in space. Here $\eps$ is an arbitrarily small discretization parameter which
will be sent to 0 at the end of all calculations, i.e. $\eps\ra 0$.
For the cases considered in this manuscript, tensors $A^{\sigma}$ with $\sigma>1$
can be neglected since they will have a vanishing contribution in the limit $\eps\ra 0$. We will 
thus use the following shorthand notations interchangeably:
\begin{align}
  A^{\sigma}(x)=
    \begin{tikzpicture}[baseline = (X.base),every node/.style={scale=0.9},scale=1.0]
      \draw (0.4,2.4) node (X) {$A(x)$};
      \draw[rounded corners] (0,2.0) rectangle (0.8,2.8);
      \draw (-0.4,2.4) -- (0.0,2.4);
      \draw (0.8,2.4) -- (1.2,2.4);
      % top vertical lines 
      \draw (0.4,1.8) -- (0.4,2.0);
    \end{tikzpicture} 
    =
  \left(
    \begin{array}{c}
      1+\epsilon Q(x)\\
      \sqrt{\epsilon}R(x)
    \end{array}
  \right)=
    \begin{tikzpicture}[baseline = (X.base),every node/.style={scale=0.7},scale=1.2]
      \draw (0.4,2.4) node (X) {
        $
        \begin{array}{c}
            \mathbbm{1+}\epsilon Q(x)\\
            \sqrt{\epsilon}R(x)
          \end{array}
        $
      };
      \draw[rounded corners] (0,2.0) rectangle (0.8,2.8);
      \draw (-0.4,2.4) -- (0.0,2.4);
      \draw (0.8,2.4) -- (1.2,2.4);
      % top vertical lines 
      \draw (0.4,1.8) -- (0.4,2.0);
    \end{tikzpicture} \nonumber
\end{align}
where we have suppressed contributions with $\sigma>1$. In the translational 
invariant case $A^{\sigma}(x)=A^{\sigma}=const.$, the cMPS can be drawn as
\begin{align*}
\ket{\Psi} =\dots
\begin{tikzpicture}[baseline = (X.base),every node/.style={scale=1.2},scale=0.75]
\draw[rounded corners] (1.0,1.0) rectangle (2.0,2.0);
\draw[rounded corners] (2.4,1) rectangle (3.4,2);
%VERTICAL LINES
\draw (2.9,1.0) -- (2.9,0.5); 
\draw (1.5,1.0) -- (1.5,0.5); 
%HORIZONTAL LINES
\draw (0.6,1.5) -- (1.0,1.5); 
\draw (2.0,1.5) -- (2.4,1.5); 
\draw (3.4,1.5) -- (3.8,1.5); 
\draw (1.5,1.5) node (X) {$A$};
\draw (2.9,1.5) node (X) {$A$};
\end{tikzpicture}\dots\;.
\end{align*}

\subsection{Regauging techniques for cMPS}

Two objects of central importance in any MPS calculation in the thermodynamic limit
are the left and right reduced steady-state density matrices
$\bra{l(x)}, \ket{r(x)}\in \mathbbm{C}^{D\times D}$.
They are defined as the eigenmatrices to eigenvalue $\eta=1$ of the so-called unit-cell
transfer operator $\mathcal{T}(0,L)$
\begin{align}\label{eq:transop}
  &\mathcal{T}(0,L)=\mathcal{P}e^{\int_0^LT(x)dx}\nonumber\\
  &T(x)=Q(x)\otimes\id+\id\otimes Q^*(x) + R(x)\otimes R^*(x). 
\end{align}
A star $^*$ denotes complex conjugation.
In formulas, $\bra{l(x)}$ and $\ket{r(x)}$ are given by
\begin{align}
  \bra{l(0)}\mathcal{T}(0,L)=\bra{l(0)}\label{eq:reducedsteadystate1}\\
  \mathcal{T}(0,L)\ket{r(L)}=\ket{r(L)}\label{eq:reducedsteadystate2}\\
  \bra{l(x)}=\bra{l(0)}\mathcal{T}(0,x)\label{eq:reducedsteadystate3}\\
  \ket{r(x)}=\mathcal{T}(x,L)\ket{r(L)}\label{eq:reducedsteadystate4}.
\end{align}
The bra-ket notation here should emphasizes the vector-character of $\bra{l(x)}$ and $\ket{r(x)}$.
We use the following diagrams to represent generic
matrices $\bra{f(x)}$ and $\ket{g(x)}$ of this character:
\begin{align}
f(x)=\bra{f(x)}=
\begin{tikzpicture}[baseline = (X.base),every node/.style={scale=1.0},scale=1.0]
    \draw (-0.35,0.2) to [out=90,in=180] (-0.15,0.5);
    \draw (-0.35,-0.2) to [out=270,in=180] (-0.15,-0.5);
    \draw[rounded corners] (-0.75,-0.22) rectangle (0.1,0.22);
    \draw (-0.35,0.0) node (X) {
      $f(x)$
    };
  \end{tikzpicture} \nonumber\;\;\;\\
  g(x)=\ket{g(x)}=
  \begin{tikzpicture}[baseline = (X.base),every node/.style={scale=1.0},scale=1.0]
    \draw (-0.35,0.2) to [out=90,in=0] (-0.55,0.5);
    \draw (-0.35,-0.2) to [out=270,in=0] (-0.55,-0.5);
    \draw[rounded corners] (-0.75,-0.22) rectangle (0.1,0.22);
    \draw (-0.35,0.0) node (X) {
      $g(x)$
    };
  \end{tikzpicture} \nonumber\;\;.
\end{align}
The cMPS transfer operator $T(x)$ of \Eq{eq:transop} acts as the derivative operator of
$\ket{r(x)},\bra{l(x)}$,
\begin{align}
  \frac{d\ket{r(x)}}{dx}&=T\ket{r(x)}\label{eq:transopoder}\\&=Q(x)r(x)+r(x)Q^{\dagger}(x)+R(x)r(x)R^{\dagger}(x)\nonumber\\
  \frac{d\bra{l(x)}}{dx}&=\bra{l(x)}T\label{eq:transopodel}\\&=l(x)Q(x)+Q^{\dagger}(x)l(x)+R^{\dagger}(x)l(x)R(x)\nonumber.
\end{align}
To obtain $\bra{l(x)}$ (or $\ket{r(x)}$) for a given $\bra{l(0)}$ (or $\ket{r(L)}$) at any point $x>0$ 
inside the unit-cell one has to solve the boundary value problem for the system of ordinary 
differential equations (ODEs) of \Eq{eq:transopoder} and \Eq{eq:transopodel}.
In this paper we use a fifth order Dormand-Prince \cite{dormand_family_1980} routine 
as provided by the {\it scientific python} package (dopri5) to integrate these equations.
The dopri5 routine (as many other high-accuracy solvers for ODEs)
chooses the step size in the integration according to a specified error bound.
Hence, it is crucial that the differential 
operator $T(x)$ can be evaluated at arbitrary points
$x\in I$. Using the above introduced diagrams,
the evolution of $\bra{l(0)}$ over the unit-cell can be drawn as
an infinite tensor network of the form 
\begin{equation}
    \begin{tikzpicture}[baseline = (X.base),every node/.style={scale=0.90},scale=0.85]
      \draw (-3.7,0.9) node (X) {$\bra{l(0)}\mathcal{T}(0,L)=$};
      \draw (3.7,0.9) node (X) {.};
      \draw (-1.95,1.14) to [out=90,in=180] (-1.2,1.44);
      \draw (-1.95,0.7) to [out=270,in=180] (-1.2,0.4);
      \draw[rounded corners] (-2.3,0.7) rectangle (-1.45,1.14);
      \draw (-1.9,0.92) node (X) {
        $l(0)$
      };

      \draw (-0.8,0.4) node (X) {$A(0)$};
      \draw[rounded corners] (-1.2,0) rectangle (-0.4,0.8);
      
      \draw (0.4,0.4) node (X) {$A(\eps)$};
      \draw[rounded corners] (0,0) rectangle (0.8,0.8);

      \draw (-0.8,1.4) node (X) {$A(0)$};
      \draw (0.4,1.4) node (X) {$A(\eps)$};
      
      \draw[rounded corners] (-1.2,1.0) rectangle (-0.4,1.8);
      \draw[rounded corners] (0,1.0) rectangle (0.8,1.8);

      % center horizontal lines 

      \draw (-0.4,1.4) -- (0.0,1.4);
      \draw (0.8,1.4) -- (1.2,1.4);
      
      % bottom horizontal lines 

      \draw (-0.4,0.4) -- (0.0,0.4);
      \draw (0.8,0.4) -- (1.2,0.4);
      
      % bottom vertical lines 
      \draw (0.4,0.8) -- (0.4,1.0);
      \draw (-0.8,0.8) -- (-0.8,1.0);

      \draw (1.5,0.9) node (X) {$\cdots$};

      % bottom row of rectangles
      \draw (2.6,0.4) node (X) {$A(L)$};
      \draw[rounded corners] (2.2,0) rectangle (3,0.8);

      % top row of rectangles
      \draw (2.6,1.4) node (X) {$A(L)$};
      \draw[rounded corners] (2.2,1.0) rectangle (3.0,1.8);

      % center horizontal lines 
      \draw (3,1.4) -- (3.4,1.4);
      \draw (1.8,1.4) -- (2.2,1.4);

      % bottom horizontal lines 
      \draw (3,0.4) -- (3.4,0.4);
      \draw (1.8,0.4) -- (2.2,0.4);
      
      % bottom vertical lines 
      \draw (2.6,0.8) -- (2.6,1.0);
      \end{tikzpicture}
\end{equation}

One of the main obstacles in cMPS optimization is the lack of algorithms to change the so called 
{\it gauge} \cite{haegeman_calculus_2013} of a generic cMPS. Algorithms for optimization of
lattice MPS make frequent use
of such regauging techniques \cite{schollwock_density-matrix_2011,verstraete_matrix_2009}
in order to improve stability and speed up convergence. 
For the simplest case of a translational invariant (c)MPS $\ket{\Psi}$ with tensors $A^{\sigma}$,
a gauge transformation is a similarity transformation 
\be
A^{\sigma}\la XA^{\sigma}X^{-1}
\ee
with an arbitrary invertible matrix $X$.
Such a gauge transformation can be used to bring the (c)MPS
tensors into left or right canonical form \cite{mcculloch_density-matrix_2007}
\begin{equation}
  \begin{split}
    \begin{tikzpicture}[baseline = (X.base),every node/.style={scale=1.0},scale=0.8]
      \draw (-0.8,0.4) node (X) {$A$};
      \draw[rounded corners] (-1.2,0) rectangle (-0.4,0.8);
      \draw (-0.8,1.4) node (X) {$A$};
      \draw (0.4,0.9) node (X) {$=$};
      \draw (1.9,0.9) node (X) {\quad\quad\quad(left)};
      \draw[rounded corners] (-1.2,1.0) rectangle (-0.4,1.8);
      \draw (-1.2,0.4) to [out=180,in=180] (-1.2,1.4) ;
      \draw (1.2,0.4) to [out=180,in=180] (1.2,1.4) ;
      % center horizontal lines 
      \draw (-0.4,1.4) -- (0.0,1.4);
      % bottom horizontal lines 
      \draw (-0.4,0.4) -- (0.0,0.4);
      % bottom vertical lines 
      \draw (-0.8,0.8) -- (-0.8,1.0);
      \end{tikzpicture} \\
    \begin{tikzpicture}[baseline = (X.base),every node/.style={scale=1.0},scale=0.8]
      \draw (-0.8,0.4) node (X) {$A$};
      \draw[rounded corners] (-1.2,0) rectangle (-0.4,0.8);
      \draw (-0.8,1.4) node (X) {$A$};
      \draw (0.4,0.9) node (X) {$=$};
      \draw (1.73,0.9) node (X) {\quad\quad\quad(right).};
      \draw[rounded corners] (-1.2,1.0) rectangle (-0.4,1.8);
      \draw (-0.4,0.4) to [out=0,in=0] (-0.4,1.4) ;
      \draw (0.95,0.4) to [out=0,in=0] (0.95,1.4) ;
      % center horizontal lines 
      \draw (-1.6,1.4) -- (-1.2,1.4);
      % bottom horizontal lines 
      \draw (-1.6,0.4) -- (-1.2,0.4);
      % bottom vertical lines 
      \draw (-0.8,0.8) -- (-0.8,1.0);
      \end{tikzpicture} 
    \end{split}
\end{equation}
\raisebox{-4pt}{\tikz{\draw (0.0,0.0) to [out=0,in=0] (0.0,0.5);}} and \raisebox{-4pt}{\tikz{\draw (0.0,0.0) to [out=180,in=180] (0.0,0.5);}} are graphical notation for identity
operators. For non-translational invariant lattice MPS,
regauging techniques make frequent use of singular value (SV) and QR
decomposition. For non-translational invariant cMPS 
on the other hand, SV and QR decomposition have not yes been developed, and hence no methods for
regauging such a cMPS are available so far. In the following we propose a method which fills
this gap and allows for an arbitrary regauging of a periodic cMPS in the thermodynamic limit. 
Let us first recall the definition of left and right orthogonality of a cMPS 
\cite{haegeman_calculus_2013}. Similar to the case of lattice MPS 
\cite{mcculloch_density-matrix_2007}, cMPS matrices $Q_{l}(x), R_{l}(x)$ or $Q_{r}(x), R_{r}(x)$
are said to be in left or right orthogonal form at position $x$ if they obey
\begin{align}
  \bra{\mathbbm{1}}T_l(x)&=Q_l(x)+Q_l^{\dagger}(x)+R_l^{\dagger}(x)R_l(x)=0\label{eq:leftortho}\\
  T_r(x)\ket{\mathbbm{1}}&=Q_r(x)+Q_r^{\dagger}(x)+R_r(x)R_r^{\dagger}(x)=0\label{eq:rightortho}.
\end{align}
The methods we are presenting in the following are generalizations of 
normalization and regauging procedures for periodic lattice MPS to the case
of cMPS.

{\bf \underline{Normalization of a cMPS:}}
As the first step, we determine the steady-state reduced density matrices $\bra{l(0)},\ket{r(L)}$
by finding the dominant left and right eigenmatrices of $\mathcal{T}(0,L)$ (using for example a
sparse eigensolver, in combination with the dopri5 routine):
\begin{align}
  \bra{l(0)}\mathcal{T}(0,L)=\eta\bra{l(0)}\nonumber\\
  \mathcal{T}(0,L)\ket{r(L)}=\eta\ket{r(L)}\nonumber
\end{align}
with $\eta\in \mathbbm{R}$. The state is normalized if $\eta=1$. For $\eta\neq 1$, we can normalize it by the transformation
\begin{align}
  Q(x)\la Q(x)-\frac{\ln\eta}{2L}\mathbbm{1}
\end{align}
at any position $x$. 

{\bf \underline{Left canonical form of a cMPS:}}
From $\bra{l(0)}$ we then obtain $\bra{l(x)}$ from
\Eq{eq:reducedsteadystate3}, which is used to transform
the cMPS matrices into left orthogonal form by the gauge transformation 
\begin{align}\label{eq:gaugedcmps}
\left(
  \begin{array}{c}
    1+\epsilon Q(x)\\
    \sqrt{\epsilon}R(x)
  \end{array}
  \right)
\rightarrow
\sqrt{l(x)}
\left(
  \begin{array}{c}
    1+\epsilon Q(x)\\
    \sqrt{\epsilon}R(x)
  \end{array}
  \right)
  \Big[\sqrt{l(x+\epsilon)}\Big]^{-1}
\end{align}
where $\sqrt{l(x)}$ is the matrix square root of $l(x)$. We have simply 
inserted an identity $\mathbbm{1}=\sqrt{l(x)}[\sqrt{l(x)}]^{-1}$ on each link between the tensors
$\left(
  \begin{array}{c}
    1+\epsilon Q(x)\\
    \sqrt{\epsilon}R(x)
  \end{array}
  \right).$
We now Taylor-expand $[\sqrt{l(x+\eps)}]^{-1}$ 
in order to express \Eq{eq:gaugedcmps} with objects at position $x$. 
Expanding $\sqrt{l(x+\eps)}$ into
\begin{align}
    \sqrt{l(x+\epsilon)}=\sqrt{l(x)}+\epsilon \frac{d\sqrt{l(x)}}{dx}\nonumber\\
    =\Big(1+\epsilon \underbrace{\frac{d\sqrt{l(x)}}{dx}\Big[\sqrt{l(x)}\Big]^{-1}}_{\phi(x)}\Big)\sqrt{l(x)}\nonumber
\end{align}
we obtain
\begin{align}
    \Big[\sqrt{l(x+\epsilon)}\Big]^{-1}=\Big[\sqrt{l(x)}\Big]^{-1}\Big(1-\epsilon\phi(x)\Big)+\mathcal{O}(\epsilon^2).\nonumber
\end{align}
If we insert this back into \Eq{eq:gaugedcmps} we get the result
\begin{align*}
\sqrt{l(x)}
\left(
  \begin{array}{c}
    1+\epsilon Q(x)\\
    \sqrt{\epsilon}R(x)
  \end{array}
  \right)
  \Big[\sqrt{l(x+\epsilon)}\Big]^{-1}=\\
  \left(
  \begin{array}{c}
    1+\epsilon \big(\sqrt{l(x)}Q(x)\Big[\sqrt{l(x)}\Big]^{-1}-\phi(x)\big)\\
    \sqrt{\epsilon}\sqrt{l(x)}R(x)\Big[\sqrt{l(x)}\Big]^{-1}
  \end{array}
  \right)
\end{align*}
from which we can read off the transformation that takes $Q(x),R(x)$ into their
left-orthogonal form:
\begin{align}
  Q_l(x)=&\sqrt{l(x)}Q(x)\Big[\sqrt{l(x)}\Big]^{-1}-\frac{d\sqrt{l(x)}}{dx}\Big[\sqrt{l(x)}\Big]^{-1}\\
  R_l(x)=&\sqrt{l(x)}R(x)\Big[\sqrt{l(x)}\Big]^{-1}.
\end{align}
A similar procedure can be used to obtain the right canonical form of a cMPS.

{\bf \underline{Central canonical form of a cMPS:}}
The above techniques can also be used to obtain what we call the
{\it central canonical form} of an inhomogeneous cMPS. 
We will only state the result here and refer the reader to the appendix for a detailed calculation. The central 
canonical form of an inhomogeneous cMPS is given by matrix functions $\Gamma_Q(x),\Gamma_R(x),C(x)$ and $\frac{dC(x)}{dx}$
such that the left and right orthogonal matrices $Q_l(x),R_l(x)$ and $Q_r(x),R_r(x)$ are obtained from
\begin{align}
    Q_l(x)&=C(x)\Gamma_Q(x)\nonumber\\
    R_l(x)&=C(x)\Gamma_R(x)\nonumber\\
    Q_r(x)&=\Gamma_Q(x)C(x)+[C(x)]^{-1}\frac{dC}{dx}\nonumber\\
    R_r(x)&=\Gamma_R(x)C(x)\nonumber.
\end{align}
It is a simple matter to verify that the matrices
\begin{align}
  \Gamma_Q(x)&=\frac{1}{\sqrt{r(x)}}Q(x)\frac{1}{\sqrt{l(x)}}\\
  &-\frac{1}{\sqrt{r(x)}}\frac{1}{\sqrt{l(x)}}\frac{d\sqrt{l(x)}}{dx}\frac{1}{\sqrt{l(x)}}\nonumber\\
  \Gamma_R(x)&=\frac{1}{\sqrt{r(x)}}R(x)\frac{1}{\sqrt{l(x)}}\\
  C(x)&=\sqrt{l(x)}\sqrt{r(x)}
\end{align}
will yield the desired result (see \Eq{eq:centralcanonicalform} for a diagrammatic representation of
a cMPS in central canonical form).
Note that unlike the central canonical form for homogeneous cMPS, 
we do not use an SVD to decompose $C(x)$ into its singular values.

The center, left and right orthogonal tensors $Q_C(x),R_C(x),Q_l(x),R_l(x)$ and $Q_r(x),R_r(x)$ are
related by the transformation
\begin{align}
  &Q_C(x)\equiv Q_l(x)C(x)\nonumber\\
  &C(x)Q_r(x)=Q_C(x)+\frac{dC}{dx}\nonumber\\
  &R_C(x)\equiv R_l(x)C(x)=C(x)R_r(x)\nonumber
\end{align}

\subsection{Regauging for spMPS}
So far the discussion has been general, without any reference to a particular
representation of a cMPS. For the case of spMPS, the transformations can be
implemented by making use of b-spline interpolations.
The normalization of a spMPS is carried out by choosing a set 
of (not necessarily) equidistantly spaced points $x_i\in [0,L]$
and transforming the matrices $Q(x_i)$ at these points according to
\begin{align}
  Q(x_i)\la Q(x_i)-\frac{\ln\eta}{2L}\mathbbm{1}.
\end{align}
A normalized spMPS is obtained from an interpolation of the points $(x_i,Q(x_i))$
(note that $R(x)$ remains unchanged). In the following we use $x_i$ with subscript $i=1,\dots,P$ 
to label the interpolation points of the cMPS matrices.
To implement the regauging, we obtain a numerical solution  $l(x_n)$ of 
\Eq{eq:reducedsteadystate3} on a discrete set of points $x_n\in [0,L],n=1\dots N$
(we use $x_n$ with subscript $n=1,\dots N$ to denote interpolation points
of $l(x_n),r(x_n)$ or other objects of the same character). 
A smooth function $l(x)$ is then again obtained from
interpolating the points $(x_n,l(x_n))$, as described above. A similar approach
is applied to obtain a smooth function $\sqrt{l(x)}$.
The quality of the left and right orthogonality depends
on the number of grid points $N$ and the order $k$ used to obtain the b-spline functions $l(x)$
and $\sqrt{l(x)}$. For the cases considered in this paper, we empirically found that using $P\approx 50$,
$N=200$ points $x_n$ and interpolating polynomials of order $k=5$, an accuracy
$\lVert \bra{\mathbbm{1}} T_l(x)\rVert<\mathcal{O}(10^{-8})$
(and similar for right orthogonalization) is easily achievable.

\section{Optimization of spMPS}
In this section we will detail a method to obtain an approximate ground state of 
a quantum field theory that possesses a non-trivial unit-cell periodicity $L$. 
We will first summarize the basic strategy and then move on to explain the individual steps
in more detail. 

\begin{enumerate}
\item Given a spMPS $Q(x), R(x)$, choose a set of points $\{x_i\}$.
\item 
  Regauge the state into the central canonical form with respect to the points $\{x_i\}$, 
  i.e. calculate $Q_C(x_i), R_C(x_i),Q_l(x_i), R_l(x_i),Q_r(x_i), R_r(x_i),C(x_i)$
  (and possibly their derivatives, see below).
\item Calculate an update $V(x_i),W(x_i)$ to the matrices $Q_C(x_i),R_C(x_i)$ (see below),
  \begin{align}
    Q_C(x_i)\ra \tilde Q(x_i)=Q_C(x_i)-\alpha V(x_i)\\
    R_C(x_i)\ra \tilde R(x_i)=R_C(x_i)-\alpha W(x_i).
  \end{align}
  This is expected to lower the energy of the state. $\alpha$ is a suitably chosen 
  small, real number.
\item 
  Obtain the matrices 
  \begin{align}
    Q(x_i)=\tilde Q(x_i)[C(x_i)]^{-1},\\
    R(x_i)=\tilde R(x_i)[C(x_i)]^{-1}
  \end{align}
\item Use a b-spline interpolation on the points 
  $(x_i,Q(x_i)), (x_i,R(x_i))$ to obtain a new spMPS $Q(x),R(x)$
  and go back to 2. 
\\
\end{enumerate}
We will demonstrate in the following our proposed method for 
a non-integrable gas of Lieb-Liniger \cite{lieb_exact_1963-1,lieb_exact_1963}
bosons in a periodic potential. The Hamiltonian is given by
\begin{align}
  H &= \int dx\, h(x)\equiv\nonumber\\
  &\int dx\, \Big(\frac{1}{2m} \partial_x\psi^\dagger(x) \partial_x\psi(x) + 
  \mu(x) \psi^\dagger(x) \psi(x)\nonumber\\
  &+ g \,\psi^\dagger(x) \psi^\dagger(x) \psi(x) \psi(x)\Big),\label{eq:Ham}\\
  \mu(x)&=\mu_0\Big(\cos(\frac{2\pi x}{L})-1\Big)^2-\frac{1}{2}.\label{eq:chempot}
\end{align}

The algorithm sketched above is a generalization of the algorithm presented in 
\cite{ganahl_continuous_2017} to 
the case of a non-homogeneous cMPS. 
We will here briefly recall the main steps of the method proposed in 
\cite{ganahl_continuous_2017} .
In the translational invariant case, the cMPS can be parametrized by two constant matrices $Q,R$,
which can be gauged into left-orthogonal form $Q_l,R_l$ (see \Eq{eq:leftortho}).
Pictorially, the state is given by
\begin{align*}
\ket{\Psi} =\dots
\begin{tikzpicture}[baseline = (X.base),every node/.style={scale=0.75},scale=.85]
\draw[rounded corners] (1.0,1.0) rectangle (2.0,2.0);
\draw[rounded corners] (2.4,1) rectangle (3.4,2);
%VERTICAL LINES
\draw (2.9,1.0) -- (2.9,0.5); 
\draw (1.5,1.0) -- (1.5,0.5); 
%HORIZONTAL LINES
\draw (0.6,1.5) -- (1.0,1.5); 
\draw (2.0,1.5) -- (2.4,1.5); 
\draw (3.4,1.5) -- (3.8,1.5); 
\draw (1.5,1.5) node (X) {
$\begin{array}{c}
  \mathbbm{1+}\epsilon Q\\
  \sqrt{\epsilon}R
\end{array}
$};
\draw (2.9,1.5) node (X) {
$\begin{array}{c}
  \mathbbm{1+}\epsilon Q\\
  \sqrt{\epsilon}R
\end{array}
$};
\end{tikzpicture}\dots = \dots
\begin{tikzpicture}[baseline = (X.base),every node/.style={scale=0.75},scale=.85]
%SECOND STa
\draw[rounded corners] (1.0,1.0) rectangle (2.0,2.0);
\draw[rounded corners] (2.4,1) rectangle (3.4,2);
%VERTICAL LINES
\draw (2.9,1.0) -- (2.9,0.5); 
\draw (1.5,1.0) -- (1.5,0.5); 
%HORIZONTAL LINES
\draw (0.6,1.5) -- (1.0,1.5); 
\draw (2.0,1.5) -- (2.4,1.5); 
\draw (3.4,1.5) -- (3.8,1.5); 
\draw (1.5,1.5) node (X) {
$\begin{array}{c}
  \mathbbm{1+}\epsilon Q_l\\
  \sqrt{\epsilon}R_l
\end{array}
$};
\draw (2.9,1.5) node (X) {
$\begin{array}{c}
  \mathbbm{1+}\epsilon Q_l\\
  \sqrt{\epsilon}R_l
\end{array}
$
};
\end{tikzpicture} \dots .
\end{align*}
The optimization for the homogeneous case proceeds by gauging the state into the central 
canonical form
\begin{align*}
\ket{\Psi} =\dots
\begin{tikzpicture}[baseline = (X.base),every node/.style={scale=0.70},scale=0.9]
%BOXES
\draw[rounded corners] (1.0,1.0) rectangle (2.0,2.0);
\draw[rounded corners] (2.8,1.5) circle (0.35);
\draw[rounded corners] (3.6,1) rectangle (4.6,2);
\draw[rounded corners] (5.4,1.5) circle (0.35);
\draw[rounded corners] (6.2,1.0) rectangle (7.2,2.0);
%HORIZONTAL LINES
\draw (0.6,1.5) -- (1.0,1.5); 
\draw (2.0,1.5) -- (2.45,1.5); 
\draw (3.15,1.5) -- (3.6,1.5); 
\draw (4.6,1.5) -- (5.05,1.5); 
\draw (5.75,1.5) -- (6.2,1.5);
\draw (7.2,1.5) -- (7.6,1.5); 
%VERTICAL LINES
\draw (1.5,1.0) -- (1.5,0.5);
\draw (4.1,1.0) -- (4.1,0.5);
\draw (6.7,1.0) -- (6.7,0.5);  
%TEXT/MATH
\draw (1.5,1.5) node (X) {
$\begin{array}{c}
  \lambda+\epsilon Q_C\\
  \sqrt{\epsilon}R_C
\end{array}
$};
\draw (2.8,1.5) node (X) {
$\lambda^{-1}$};
\draw (5.4,1.5) node (X) {
$\lambda^{-1}$};
\draw (4.1,1.5) node (X) {
$\begin{array}{c}
 \lambda +\epsilon Q_C\\
  \sqrt{\epsilon}R_C
\end{array}
$
};
\draw (6.7,1.5) node (X) {
$\begin{array}{c}
 \lambda+\epsilon Q_C\\
  \sqrt{\epsilon}R_C
\end{array}
$};
\end{tikzpicture} \dots ,
\end{align*}
were $\lambda$ is a diagonal matrix containing the Schmidt values, and 
\begin{align}
  Q_C=Q_l\lambda\\
  R_C=R_l\lambda.
\end{align}
A local update $V,W$ to the matrices $Q_C,R_C$ is then calculated from the gradient of the energy,
\begin{align}
    V,W=\underset{Q_C,R_C}{\text{argmin}}\frac{\braket{\Psi|H|\Psi}}{\braket{\Psi|\Psi}},
\end{align}
(see appendix in \cite{ganahl_continuous_2017}) producing a new set of tensors
\begin{align*}
  \tilde Q=(Q_C-\alpha V)\lambda^{-1}\\
  \tilde R=(R_C-\alpha W)\lambda^{-1}
\end{align*}
from which the above described procedure can be restarted ($\alpha$ is a small, real number).

Our generalization of this method to an inhomogeneous setting starts by
choosing a set of (not necessarily) equidistantly spaced points ${x_i\in}[0,L], {{i=}1\dots P}$, 
and bringing a spMPS into the central canonical form around these points. 
Pictorially, the state decomposition is given by
\begin{equation}
  \dots
  \begin{tikzpicture}[baseline = (X.base),every node/.style={scale=0.55},scale=.65]
    \draw[rounded corners] (0.5,1.0) rectangle (2.0,2.0);
    \draw[rounded corners] (2.3,1.3) rectangle (3.6,1.7);
    \draw[rounded corners] (3.9,1.3) rectangle (5.5,1.7);
    % vertical lines
    \draw (1.25,1.0) -- (1.25,0.6);
    % horizontal lines
    \draw (0.2,1.5) -- (0.5,1.5);
    \draw (2.0,1.5) -- (2.3,1.5);
    \draw (3.6,1.5) -- (3.9,1.5);
    \draw (1.25,1.5) node (X) {
      $\begin{array}{c}
        {\mathbbm{1}+}\epsilon Q_l(x_i)\\
        \sqrt{\epsilon}R_l(x_i)
      \end{array}
      $};
    \draw (2.9,1.5) node (X) {$C({x_i+}\eps)$};
    \draw (4.7,1.5) node (X) {$C^{-1}({x_i+}\eps)$};
  \end{tikzpicture} \dots 
  \begin{tikzpicture}[baseline = (X.base),every node/.style={scale=0.55},scale=0.8]
    \draw[rounded corners] (0.5,1.0) rectangle (2.0,2.0);
    \draw[rounded corners] (2.3,1.3) rectangle (3.6,1.7);
    \draw[rounded corners] (3.9,1.3) rectangle (5.5,1.7);
    % vertical lines
    \draw (1.25,1.0) -- (1.25,0.6);
    % horizontal lines
    \draw (0.2,1.5) -- (0.5,1.5);
    \draw (2.0,1.5) -- (2.3,1.5);
    \draw (3.6,1.5) -- (3.9,1.5);
    \draw (1.25,1.5) node (X) {
      $\begin{array}{c}
        {\mathbbm{1}+}\epsilon Q_l(x_{i+1})\\
        \sqrt{\epsilon}R_l(x_{i+1})
      \end{array}
      $};
    \draw (2.9,1.5) node (X) {$C({x_{i+1}+}\eps)$};
    \draw (4.7,1.5) node (X) {$C^{-1}({x_{i+1}+}\eps)$};
  \end{tikzpicture} \dots 
\end{equation}
which to order $\mathcal{O}(\eps^{3/2})$ is equivalent to 
\begin{widetext}
  \begin{equation}\label{eq:centralcanonicalform}
    \begin{split}
      \dots
      \begin{tikzpicture}[baseline = (X.base),every node/.style={scale=0.65},scale=0.75]
        \draw[rounded corners] (1.5,1.0) rectangle (5.3,2.0);
        \draw[rounded corners] (5.6,1.3) rectangle (7.2,1.7);
        % vertical lines
        \draw (3.4,1.0) -- (3.4,0.6);
        % horizontal lines
        \draw (1.2,1.5) -- (1.5,1.5);
        \draw (5.3,1.5) -- (5.6,1.5);
        \draw (7.2,1.5) -- (7.5,1.5);
        \draw (3.4,1.5) node (X) {
          $\begin{array}{c}
            {C(x_i)+}\epsilon (Q_C(x_i)+C'(x_i))\\
            \sqrt{\epsilon}R_C(x_i)
          \end{array}
          $};
        \draw (6.4,1.5) node (X) {$C^{-1}({x_i+}\eps)$};
      \end{tikzpicture}
      \dots
      \begin{tikzpicture}[baseline = (X.base),every node/.style={scale=0.65},scale=0.75]
        \draw[rounded corners] (1.1,1.0) rectangle (5.7,2.0);
        \draw[rounded corners] (6.0,1.3) rectangle (7.9,1.7);
        % vertical lines
        \draw (3.4,1.0) -- (3.4,0.6);
        % horizontal lines
        \draw (0.8,1.5) -- (1.1,1.5);
        \draw (5.7,1.5) -- (6,1.5);
        \draw (7.9,1.5) -- (8.2,1.5);
        \draw (3.4,1.5) node (X) {
          $\begin{array}{c}
            {C(x_{i+1})+}\epsilon (Q_C(x_{i+1})+C'(x_{i+1}))\\
            \sqrt{\epsilon}R_C(x_{i+1})
          \end{array}
          $};
        \draw (6.9,1.5) node (X) {$C^{-1}({x_{i+1}+}\eps)$};
      \end{tikzpicture}\dots \;\;.
    \end{split}
  \end{equation}
\end{widetext}
Primes are shorthand for derivatives $\frac{d}{dx}$.
The next step is the determination of an update $V(x_i),W(x_i)$ for the matrices
$Q_C(x_i), R_C(x_i)$ at each point $x_i$. As in the homogeneous case, 
we choose as update the local gradient 
\begin{align}
  V(x_i)=\frac{\delta}{\delta Q_C^*(x_i)}\frac{\braket{\Psi|H|\Psi}}{\braket{\Psi|\Psi}}\\
  W(x_i)=\frac{\delta}{\delta R_C^*(x_i)}\frac{\braket{\Psi|H|\Psi}}{\braket{\Psi|\Psi}}.
\end{align}
For the Hamiltonian \Eq{eq:Ham}, the update $W(x_i)$ to $R_C(x_i)$ has contributions
from kinetic energy, potential energy, interaction energy
and the environment (see below), 
respectively. The update $V(x_i)$ to $Q_C(x_i)$ on the other hand has 
only contributions from kinetic energy and environment. The updates decompose
into a sum of the form
\begin{align}\label{eq:update0}
  W(x_i)&=W_{kin}(x_i)+W_{pot}(x_i)+W_{int}(x_i)+W_{env}(x_i)\nonumber\\
  V(x_i)&=V_{kin}(x_i)+V_{env}(x_i).
\end{align}
A straight forward but lengthy
calculation gives the following results for the contributions to $W(x_i)$ and $V(x_i)$:
\begin{widetext}
  \begin{equation}\label{eq:update}
    \begin{split}
      W_{env}(x_i)=&H_l(x_i)R_l(x_i)C(x_i)+R_l(x_i)C(x_i)H_r(x_i)\\
      W_{pot}(x_i)=&\mu(x_i)R_l(x_i)C(x_i)\\
      W_{int}(x_i)=&g\Big(R_l^{\dagger}(x_i)R_l(x_i)R_l(x_i)C(x_i)+
      C(x_i)R_r(x_i)R_r(x_i)R_r^{\dagger}(x_i)|1)\Big)\\
      W_{kin}(x_i)=&
      \frac{1}{2m}\Big(R_l'(x_i)Q_l(x_i)C(x_i)-Q_l'(x_i)R_l(x_i)C(x_i)
      -Q_l(x_i)R_l'(x_i)C(x_i)-Q_l(x_i)R_l(x_i)C'(x_i)\\
      &+R_l(x_i)Q_l'(x_i)C(x_i)+R_l(x_i)Q_l(x_i)C'(x_i)-R_l'(x_i)C'(x_i)-
      R_l''(x_i)C(x_i)\\
      &+Q_l^{\dagger}(x_i)\big([Q_l(x_i),R_l(x_i)]+R_l'(x_i)\big)C(x_i)
      -\big([Q_l(x_i),R_l(x_i)]+R_l'(x_i)\big)C(x_i)Q_r^{\dagger}(x_i)\Big)\\
      V_{env}(x_i)=&H_l(x_i)C(x_i)+C(x_i)H_r(x_i)\\
      V_{kin}(x_i)=&\frac{1}{2m}\Big(-R_l^{\dagger}(x_i)\big([Q_l(x_i),R_l(x_i)]+\
      R_l'(x_i)\big)C(x_i)+
      \big([Q_l(x_i),R_l(x_i)]+R_l'(x_i)\big)C(x_i)R_r^{\dagger}(x_i)\Big).
    \end{split}
  \end{equation}
\end{widetext}
Here, $H_l(x_i)$ and $H_r(x_i)$ are hermitian matrices of dimensions $D\times D$ containing
the Hamiltonian environment for point $x_i$ (see appendix for details on their calculation). 
Once the updates have been calculated, the tensors $Q_C^{[m]}(x_i)$ and $R_C^{[m]}(x_i)$ 
at iteration step $m$ are changed according to 
\begin{align}
  Q_C^{[m]}(x_i)&\ra Q^{[m+1]}(x_i)\nonumber\\
  &=[Q_C^{[m]}(x_i)-\alpha V^{[m]}(x_i)][C^{[m]}(x_i)]^{-1}\nonumber\\
  R_C^{[m]}(x_i)&\ra R^{[m+1]}(x_i)\nonumber\\
  &=[R_C^{[m]}(x_i)-\alpha W^{[m]}(x_i)][C^{[m]}(x_i)]^{-1}.\nonumber
\end{align}
$\alpha\in \mathbbm{R}$ is a small stepsize parameter (typically $\alpha=10^{-4}\dots 10^{-5}$).
From the updated tensors $Q^{[m+1]}(x_i),R^{[m+1]}(x_i)$, a new spMPS is 
obtained from interpolating the points $(x_i,Q^{[m+1]}(x_i)),(x_i,R^{[m+1]}(x_i))$,
which implements a change of the variational parameters, 
\begin{align}
  \mathcal{Q}^{i,[m]}\ra \mathcal{Q}^{i,[m+1]}\nonumber\\
  \mathcal{R}^{i,[m]}\ra \mathcal{R}^{i,[m+1]}\nonumber.
\end{align}
and finishes a single update step. The procedure is then iterated until convergence is reached. 
As convergence parameter we monitor the norm of the gradient 
\begin{align}
  \lVert \mathcal{G}(x_i)\rVert=\sqrt{\tr[V(x_i)V^{\dagger}(x_i)+W(x_i)W^{\dagger(}x_i)]}.
\end{align}
We stop the iteration once $\max[\lVert \mathcal{G}(x_i)\rVert]<10^{-4}$. In our simulations 
we found that shifting the unit-cell of the system by $L/2$ every $m_{shift}=5$ 
update steps has a stabilizing effect on the simulation.

\begin{figure}
  \includegraphics[width=1.0\columnwidth]{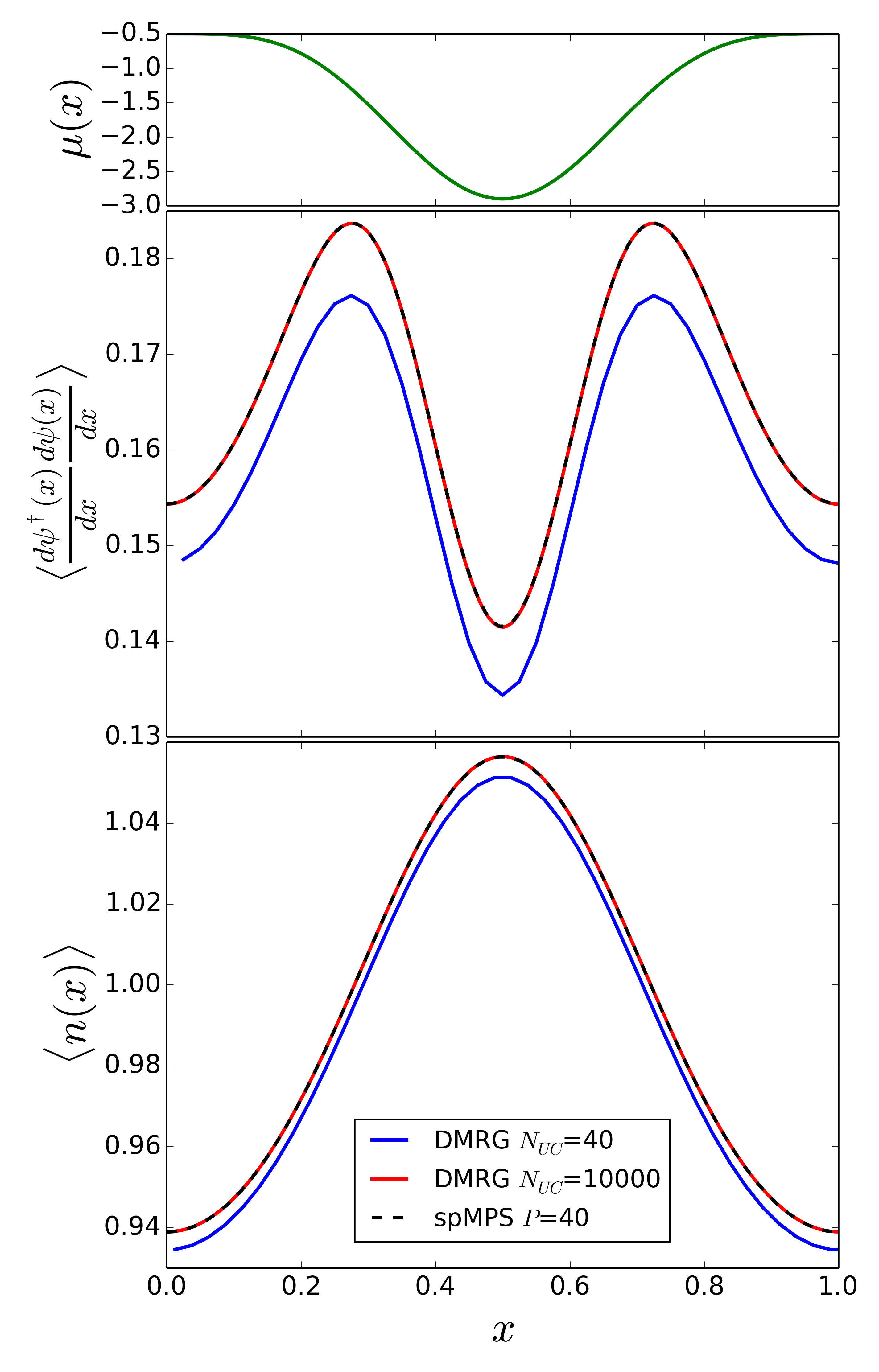}
  \caption{{\bf Observables in the ground state \Eq{eq:Ham} for $\bm{g=1.0,m=0.5,\mu_0=-0.6}$ 
      and a bond dimension $\bm{D=16}$.}
    Top panel: 
    chemical potential $\mu(x)$. We plot kinetic energy density 
    $\braket{\frac{d\psi^{\dagger}(x)}{dx}\frac{d \psi(x)}{dx}}$ (middle panel) and 
    particle density $\braket{n(x)}\equiv\braket{\psi^{\dagger}(x)\psi(x)}$(lower panel) as a 
    function of space-coordinate $x\in[0,L]$ for 
    a single unit-cell. Black dashed lines correspond to spMPS results with $P=40$
    b-spline polynomials. Blue and red solid lines are results from a lattice DMRG
    calculation, obtained from discretizing the Hamiltonian into $N_{UC}=40,10^4$
    discretization points per unit-cell, respectively.
  }\label{fig:splimpsobservables}
\end{figure}

\begin{figure}
  \includegraphics[width=1.0\columnwidth]{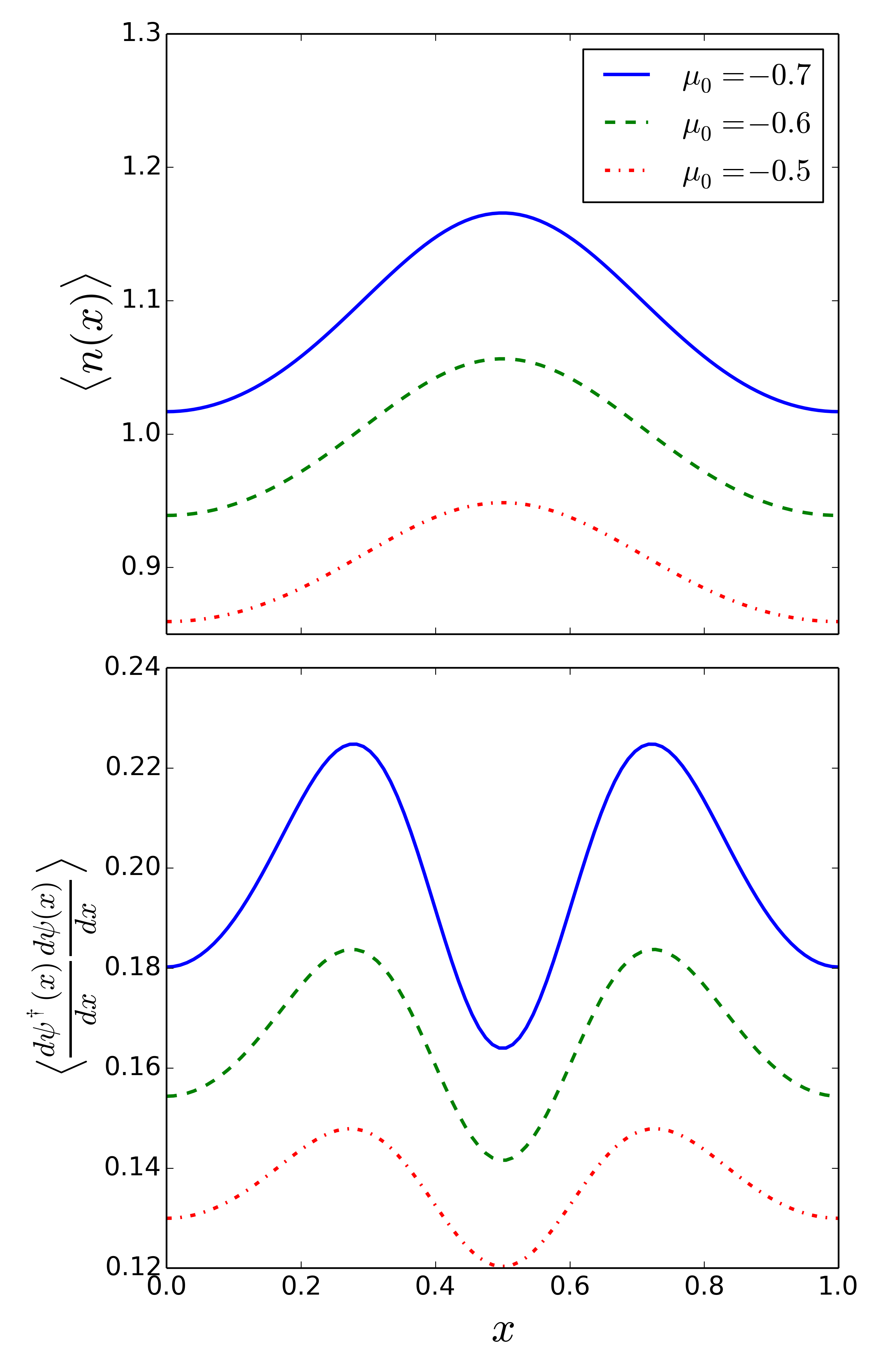}
  \caption{{\bf Ground state observables for different amplitudes of the chemical 
      potential.} Particle density $\braket{n(x)}$ (upper panel) and kinetic energy density 
    (lower panel) for different 
    amplitudes $\mu_0=-0.5,-0.6,-0.7$ and $g=1.0,m=0.5,D=16$.
  }\label{fig:splimpsobservables_2}
\end{figure} 

\section{Results and discussion}
In this last section we present results for optimized spMPS ground states for a gas of 
Lieb-Liniger bosons in a periodic potential, see \Eq{eq:Ham}. 
We emphasize at this point that a good initial state for the spMPS
optimization is important for a stable ground state optimization. The main problem we encounter 
in our simulations are diverging matrix functions $Q(x),R(x)$ with ongoing optimization. 
Usually, these divergences can have multiple causes, and it is not always possible to pin 
down the exact cause for any particular case. One well known problem is related to the step 
size $\alpha$ in the optimization and the average distance $\Delta x=x_i-x_{i-1}$ between two 
interpolation points. The so-called Courant-Levy condition determines a maximally possible step 
size $\alpha_{max}$ for a given $\Delta x$ before the iteration diverges. Additionally to this, 
we observe that for large gradients, the step size has to be taken very small in order to avoid 
any divergences. For random initial states, the gradients will typically become so large that the 
step sizes get too small to reach convergence within any reasonable 
time frame. Starting from a good initial guess for a spMPS wave function
reduces many of these problems considerably and leads to a sizeable reduction of computational run times.
In this manuscript we use an optimization and interpolation method for lattice MPS
to get a good initial guess for the spMPS optimization, which we then further
optimize with our proposed optimization method. The details of this initialization method will
be published elsewhere \cite{inprep}.

In \Fig{fig:splimpsobservables} we show results for the kinetic energy density
$\braket{\frac{d\psi^{\dagger}}{dx}\frac{d\psi}{dx}}$ (middle panel) and particle $\braket{n(x)}$
(lower panel) for an approximate ground state of \Eq{eq:Ham} as obtained by the proposed spMPS 
optimization (black dashed line), for a bond dimension $D=16$. 
The top panel shows the chemical potential $\mu(x)$ \Eq{eq:chempot} for $\mu_0=-0.6$. 
We used a spMPS of order $k=5$ and $P=40$ interpolation 
polynomials $B^k_i(x)$ within a unit-cell. The observables are exactly
periodic with the same periodicity as the chemical potential . For comparison,
we show results from a DMRG calculation, with a discretization of \Eq{eq:Ham} using $N_{UC}=40$
discretization points in the unit-cell (blue solid line). As reference, we
also show a DMRG calculation using $N_{UC}=10^4$ discretization 
points per unit-cell, which we consider to be the variational optimum for a
chosen bond dimension $D=16$, and to be practically free of discretization
artifacts. The exceptional agreement of the spMPS results with the
$N_{UC}=10^4$ DMRG calculation is evidence that the spMPS wave function is
essentially free of any discretization artifacts.

In \Fig{fig:splimpsobservables_2} we show results for particle density $\braket{n(x)}$ and kinetic energy density $\braket{\frac{d\psi^{\dagger}}{dx}\frac{d\psi}{dx}}$
for different values of the amplitude $\mu_0$ of the chemical potential (see \Eq{eq:chempot}).
As expected, raising the amplitude causes a similar increase in the amplitudes for the 
particle density, as well as the kinetic energy density.

\section{Conclusions and Outlook}
We have proposed a novel parametrization for non-translational invariant 
continuous Matrix Product States in terms of a b-spline interpolation for the 
cMPS matrix-functions $Q(x), R(x)$. 
The b-spline parametrization allows for an efficient representation of 
inhomogeneous continuous many-body wave functions,
and is free of any discretization artifacts. We extend well known regauging
techniques for lattice Matrix Product States to the case of spline-based Matrix Product States
(spMPS), and show how a recently developed optimization method for translational 
invariant cMPS can be applied to the case of inhomogeneous spMPS. As a proof-of-principle, 
we apply the method to a gas of interacting Lieb-Liniger bosons in a periodic potential,
where a comparison with DMRG calculations underpins that the proposed variational class
of spMPS is free of any discretization errors. It is thus well suited
to parametrize ground-states of continuous quantum field theories without
translational invariance. 

While these first results are very promising, much 
remains to be done. In particular, results in this paper were obtained
for a bond dimension $D=16$ which is still small compared to standard lattice cases,
where $D\sim \mathcal{O}(10^2)$ can be easily reached using dense MPS tensors $A^{\sigma}$.
The reason for these small bond dimensions in the spMPS case are related 
to the gauge degrees of freedom in cMPS. In the lattice preconditioning step,
matrix entries $Q_{\alpha\beta}(x),R_{\alpha\beta}(x)$ for small Schmidt values $\lambda_{\alpha}$
start to exhibit gauge-fluctuations on short scales. While these fluctuations
are usually small, their higher spatial derivatives quickly become large.
The spMPS update, depending on these higher derivatives, then becomes unstable.
We are currently exploring new methods for gauge-smoothing MPS states which
will help to overcome this problem, and will enable us to obtain results
for larger bond dimensions.

It is important to point out here that the above mentioned
instabilities are of a different origin than those occurring for cMPS simulations
with open boundary conditions \cite{haegeman_quantum_2015}. In the latter case,
the Schmidt spectrum of the cMPS becomes rank deficient close to the boundaries.
For simulations applying the Time Dependent Variational Principle to cMPS, this
rank deficiency leads to problems due to a necessary inversion of the Schmidt spectrum
in the algorithm. In our case, the Schmidt-spectrum is always full rank.
It has to been seen if a parametrization in terms of spMPS can overcome
such problems for the case of open boundary conditions.

The optimization method which we have proposed in this paper uses a point-wise
update of the spMPS tensors $Q(x_i),R(x_i)$ at points $x_i$, followed by 
an interpolation to new matrix-functions $\tilde Q(x),\tilde R(x)$ using the points
$(x_i,Q(x_i)),(x_i,R(x_i))$ as interpolation points. The variational parameters 
$\mathcal{Q}^i,\mathcal{R}^i$ get updated via the interpolation step.
It would be interesting to see if the tensors $\mathcal{Q}^i,\mathcal{R}^i$ 
could be updated directly as well. For example, for a given $i$, a variation of the 
tensors $\mathcal{Q}^i,\mathcal{R}^i$ introduces local changes of the tensors $Q(x),R(x)$,
where the changes are localized around position $x_i$, and are smooth in space,
due to the smoothness and locality of the b-spline polynomials $B_i^k(x)$. 
This local change gives rise to a change in the energy expectation value and,
if chosen appropriately, will lower the energy. To find an appropriate variation, one could
for example linearize the energy functional around the current 
values of $\mathcal{Q}^i,\mathcal{R}^i$. Such an approach could potentially 
greatly enhance the speed of convergence for ground state optimizations for spMPS. 

Another important direction is the extension of spMPS to systems with multiple particle species.
In this case, each species $\alpha$ gives rise to a matrix $R_{\alpha}(x)$. The b-spline
interpolation can be applied to each $R_{\alpha}(x)$ individually. To ensure the necessary
regularity conditions \cite{haegeman_calculus_2013}, suitable parametrizations for the 
matrices $R_{\alpha}(x)$ can be combined with the b-spline interpolation.

Finally, the ability to use a continuous-space parametrization of 
the many-body wave function $\ket{\Psi}$ opens the possibility to combine tools 
for solving partial differential equations (PDEs) developed in numerical mathematics
to be combined with non-perturbative tensor network approaches to interacting quantum 
field theories. Vice versa, the methods developed here may have important 
applications in the field of numerical mathematics of coupled, non-linear PDE.

\section{Acknowledgements}
The author thanks J. Rinc\'on, A. Milsted and G. Vidal for useful
discussions and G. Vidal for valuable comments on the manuscript.
The author also acknowledges support by the Simons Foundation (Many Electron Collaboration). 
Computations were made on the supercomputer Mammouth parall\`ele II from University of 
Sherbrooke, managed by Calcul Qu\'ebec and Compute Canada. 
The operation of this supercomputer is funded by the Canada Foundation for Innovation (CFI), 
the minist\`ere de l'\'Economie, de la science et de l'innovation du Qu\'ebec (MESI) and the 
Fonds de recherche du Qu\'ebec - Nature et technologies (FRQ-NT). 
This research was supported in part by Perimeter Institute for Theoretical Physics. 
Research at Perimeter Institute is supported by the Government of Canada through Industry Canada 
and by the Province of Ontario through the Ministry of Economic Development \& 
Innovation.

%\bibliography{cmps_bib}

\section{Appendix}
\setcounter{equation}{0}
\setcounter{figure}{0}
\setcounter{table}{0}
\setcounter{page}{1}
\renewcommand{\theequation}{A\arabic{equation}}
\renewcommand{\thefigure}{A\arabic{figure}}
\renewcommand{\bibnumfmt}[1]{[S#1]}

\subsection{Obtaining the central canonical form of a cMPS}
The calculation for obtaining the central canonical form
proceeds largely along the same lines as for the left-orthogonal gauge. 
The first step is the determination of the left and right steady-state reduced 
density matrices $\bra{l(x)},\ket{r(x)}$. On each link between positions $x$ and $x+\eps$,
(with a tensor $A^{\sigma}(x)$ in between)
we insert resolutions of the identity of the form $[\sqrt{l(x)}\Big]^{-1}
\sqrt{l(x)}\sqrt{r(x)}\Big[\sqrt{r(x)}\Big]^{-1}$:
\begin{widetext}
\begin{align}
\left(
  \begin{array}{c}
    1+\epsilon Q(x)\\
    \sqrt{\epsilon}R(x)
  \end{array}
  \right)
=
\Big[\sqrt{l(x)}\Big]^{-1}
\sqrt{l(x)}\sqrt{r(x)}\Big[\sqrt{r(x)}\Big]^{-1}
\left(
  \begin{array}{c}
    1+\epsilon Q(x)\\
    \sqrt{\epsilon}R(x)
  \end{array}
  \right)
  \Big[\sqrt{l(x+\epsilon)}\Big]^{-1}\sqrt{l(x+\epsilon)}\sqrt{r(x+\epsilon)}\Big[\sqrt{r(x+\epsilon)}\Big]^{-1}\label{eq:centralc}.
\end{align}
\end{widetext}
With the definition 
\be
C(x)\equiv\sqrt{l(x)}\sqrt{r(x)}\nonumber
\ee
and the expansion 
\begin{align} 
     \frac{1}{\sqrt{l(x+\epsilon)}}=\frac{1}{\sqrt{l(x)}}\Big(1-\epsilon\frac{d\sqrt{l(x)}}{dx}\Big[\sqrt{l(x)}\Big]^{-1}\Big)+\mathcal{O}(\epsilon^2)\nonumber,
\end{align}
a simple calculations turns \Eq{eq:centralc}
into
\begin{widetext}
  \begin{align*}
    \left(
      \begin{array}{c}
        1+\epsilon Q(x)\\
        \sqrt{\epsilon}R(x)
      \end{array}
    \right)
    =&
    \Big[\sqrt{l(x)}\Big]^{-1}\underbrace{C(x)\frac{1}{\sqrt{r(x)}}\left(
        \begin{array}{c}
          1+\epsilon Q(x)\\
          \sqrt{\epsilon}R(x)
        \end{array}
      \right)
      \frac{1}{\sqrt{l(x)}}\Big(1-\epsilon\frac{d\sqrt{l(x)}}{dx}\frac{1}{\sqrt{l(x)}}\Big)}_{\textrm{left normalized}}\Big(C(x)+\epsilon\frac{dC}{dx}\Big)\Big[\sqrt{r(x+\eps)}\Big]^{-1}\\
    =&
    \Big[\sqrt{l(x)}\Big]^{-1}C(x)\underbrace{\frac{1}{\sqrt{r(x)}}\left(
        \begin{array}{c}
          1+\epsilon Q(x)\\
          \sqrt{\epsilon}R(x)
        \end{array}
      \right)
      \frac{1}{\sqrt{l(x)}}\Big(1-\epsilon\frac{d\sqrt{l(x)}}{dx}\frac{1}{\sqrt{l(x)}}\Big)\Big(C(x)+\epsilon\frac{dC}{dx}\Big)}_{\textrm{right normalized}}\Big[\sqrt{r(x+\eps)}\Big]^{-1},
  \end{align*}
\end{widetext}
where we have dropped terms of order $\mathcal{O}(\eps^2)$, and we have 
indicated how to contract matrices such that a left or right normalized cMPS is obtained.
The central canonical form is defined by matrices $\Gamma_Q(x),\Gamma_R(x),C(x)$ and
$\frac{dC(x)}{dx}$, such that 
\begin{align*}
    Q_l&=C(x)\Gamma_Q(x)\\
    R_l&=C(x)\Gamma_R(x)\\
    Q_r&=\Gamma_Q(x)C(x)+[C(x)]^{-1}\frac{dC}{dx}\\
    R_r&=\Gamma_R(x)C(x).
\end{align*}
It is then just a matter of collecting 
orders of $\eps$ to see that the matrices
\begin{align*}
    \Gamma_Q(x)&=\frac{1}{\sqrt{r(x)}}Q(x)\frac{1}{\sqrt{l(x)}}\\
    &-\frac{1}{\sqrt{r(x)}}\frac{1}{\sqrt{l(x)}}
    \frac{d\sqrt{l(x)}}{dx}\frac{1}{\sqrt{l(x)}}\\
    \Gamma_R(x)&=\frac{1}{\sqrt{r(x)}}R(x)\frac{1}{\sqrt{l(x)}}\\
\end{align*}
fullfill this condition.
Note that when permuting $C(x)$ past $Q_l(x)$ or $Q_r(x)$, we have to use
\begin{align}
  C(x)Q_r(x)=Q_l(x)C(x)+\frac{dC}{dx}.
\end{align}

\subsection{Calculation of environmental contributions}
In this section we will outline the calculation of the matrices $H_l(x_i), H_r(x_i)$
encountered in \Eq{eq:update}. In DMRG parlance, these hermitian matrices are the left and right 
Hamiltonian environments of a local site at position $x_i$. For example, $H_l(x_i)$ contains
all contributions $\int_{-\infty}^{x_i}h(x)dx$ to the left of $x_i$. On a more technical level, 
$H_l$ and $H_r$ are the projections of $\int_{-\infty}^{x_i}h(x)dx$ and $\int_{x_i}^{\infty}h(x)dx$
into two reduced orthonormal basis sets at position $x_i$. 
At this point it is helpful to introduce the standard MPS diagramatic technique to
visualize the following calculations. To this end we use the diagram
\begin{equation}
  \dots
  \begin{tikzpicture}[baseline = (X.base),every node/.style={scale=0.8},scale=.95]
    %center row of rectangles
    \draw[fill=white,rounded corners] (-1.2,1.0) rectangle (-0.4,1.8);
    \draw (-0.8,1.4) node (X) {$x_i-\eps$};

    \draw[fill=white,rounded corners] (0,1.0) rectangle (0.8,1.8);
    \draw (0.4,1.4) node (X) {$x_{i}$};

    \draw[fill=white,rounded corners] (1.2,1.0) rectangle (2.0,1.8);
    \draw (1.6,1.4) node (X) {$x_i+\eps$};

    %center horizontal lines 
    \draw (2.0,1.4) -- (2.4,1.4);
    \draw (-1.6,1.4) -- (-1.2,1.4);
    \draw (-0.4,1.4) -- (0.0,1.4);
    \draw (0.8,1.4) -- (1.2,1.4);

    %bottom vertical lines 
    \draw (0.4,0.8) -- (0.4,1.0);
    \draw (-0.8,0.8) -- (-0.8,1.0);
    \draw (1.6,0.8) -- (1.6,1.0);

    %top vertical lines 
    \draw (0.4,1.8) -- (0.4,2.0);
    \draw (-0.8,1.8) -- (-0.8,2.0);
    \draw (1.6,1.8) -- (1.6,2.0);
  \end{tikzpicture} 
  \dots
\end{equation}
to denote an MPO representation of a discretization of the Hamiltonian \Eq{eq:Ham}, with lattice spacing $\eps$.
Note that the following diagrams serve only as a visual aid, and network contractions are performed solely
by integration of a system of ordinary differential equations (see below).
In this diagrammatic notation $H_l(x_i)$ and $H_r(x_i)$ are given by half-infinite tensor networks of the form
\begin{equation}
  H_l(x_i)=
    \begin{tikzpicture}[baseline = (X.base),every node/.style={scale=0.70},scale=1.2]
      \draw (0,0.1) to [out=180,in=180] (0,2.1);
      \draw (-0.6,1.1) -- (-0.2,1.1);
    \end{tikzpicture}
    \dots
    \begin{tikzpicture}[baseline = (X.base),every node/.style={scale=0.60},scale=1.2]
      % bottom row of rectangles
      \draw (-0.8,0.4) node (X) {$A_{l}({x_i-}2\eps)$};
      \draw (0.4,0.4) node (X) {$A_{l}({x_i-}\eps)$};

      \draw[rounded corners] (0,0) rectangle (0.8,0.8);
      \draw[rounded corners] (-1.2,0) rectangle (-0.4,0.8);
      
      % top row of rectangles
      \draw (-0.8,2.4) node (X) {$A_{l}({x_i-}2\eps)$};
      \draw (0.4,2.4) node (X) {$A_{l}({x_i-}\eps)$};
      
      \draw[rounded corners] (-1.2,2.0) rectangle (-0.4,2.8);
      \draw[rounded corners] (0,2.0) rectangle (0.8,2.8);
      
      % center row of rectangles
      \draw[fill=white,rounded corners] (-1.2,1.0) rectangle (-0.4,1.8);
      \draw[fill=white,rounded corners] (0,1.0) rectangle (0.8,1.8);
      
      \draw (-0.8,1.4) node (X) {${x_{i}-}2\eps$};
      \draw (0.4,1.4) node (X) {${x_{i}-}\eps$};

      % center horizontal lines 
      \draw (-1.6,1.4) -- (-1.2,1.4);
      \draw (-0.4,1.4) -- (0.0,1.4);
      
      % top horizontal lines 
      \draw (-1.6,2.4) -- (-1.2,2.4);
      \draw (-0.4,2.4) -- (0.0,2.4);
      \draw (0.8,2.4) -- (1.2,2.4);
      
      % bottom horizontal lines 
      \draw (-1.6,0.4) -- (-1.2,0.4);
      \draw (-0.4,0.4) -- (0.0,0.4);
      \draw (0.8,0.4) -- (1.2,0.4);
      
      % bottom vertical lines 
      \draw (0.4,0.8) -- (0.4,1.0);
      \draw (-0.8,0.8) -- (-0.8,1.0);
      
      % top vertical lines 
      \draw (0.4,1.8) -- (0.4,2.0);
      \draw (-0.8,1.8) -- (-0.8,2.0);
      \end{tikzpicture} 
\nonumber
\end{equation}

\begin{equation}
  H_r(x_i)=\dots\nonumber
    \begin{tikzpicture}[baseline = (X.base),every node/.style={scale=0.60},scale=1.2]
      % bottom row of rectangles
      \draw (1.6,0.4) node (X) {$A_{r}({x_i+}2\eps)$};
      \draw (0.4,0.4) node (X) {$A_{r}({x_i+}\eps)$};
      
      \draw[rounded corners] (0,0) rectangle (0.8,0.8);
      \draw[rounded corners] (1.2,0) rectangle (2.0,0.8);
      
      % top row of rectangles
      \draw (0.4,2.4) node (X) {$A_{r}({x_i+}\eps)$};
      \draw (1.6,2.4) node (X) {$A_{r}({x_i+}2\eps)$};
      
      \draw[rounded corners] (0,2.0) rectangle (0.8,2.8);
      \draw[rounded corners] (1.2,2.0) rectangle (2.0,2.8);
      
      % center row of rectangles
      \draw[fill=white,rounded corners] (0,1.0) rectangle (0.8,1.8);
      \draw[fill=white,rounded corners] (1.2,1.0) rectangle (2.0,1.8);

      \draw (0.4,1.4) node (X) {${x_{i}+}\eps$};
      \draw (1.6,1.4) node (X) {${x_{i}+}2\eps$};

      % center horizontal lines 
      \draw (0.8,1.4) -- (1.2,1.4);
      \draw (2.0,1.4) -- (2.4,1.4);
      
      % top horizontal lines 
      \draw (-0.4,2.4) -- (0.0,2.4);
      \draw (0.8,2.4) -- (1.2,2.4);
      \draw (2.0,2.4) -- (2.4,2.4);
      
      % bottom horizontal lines 
      \draw (-0.4,0.4) -- (0.0,0.4);
      \draw (0.8,0.4) -- (1.2,0.4);
      \draw (2.0,0.4) -- (2.4,0.4);
      
      % bottom vertical lines 
      \draw (0.4,0.8) -- (0.4,1.0);
      \draw (1.6,0.8) -- (1.6,1.0);
      
      % top vertical lines 
      \draw (0.4,1.8) -- (0.4,2.0);
      \draw (1.6,1.8) -- (1.6,2.0);
      \end{tikzpicture} 
    \dots
    \begin{tikzpicture}[baseline = (X.base),every node/.style={scale=0.70},scale=1.2]
      \draw (2,0.4) to [out=0,in=0] (2,2.4);
      \draw (2.6,1.4) -- (2.2,1.4);
    \end{tikzpicture}.
\end{equation}
\raisebox{-4pt}{\tikz{\draw (0.0,0.0) to [out=180,in=180] (0.0,0.5);\draw (-0.15,0.25) -- (0.0,0.25);}}
and
\raisebox{-4pt}{\tikz{\draw (0.0,0.0) to [out=0,in=0] (0.0,0.5);\draw (0.0,0.25) -- (0.15,0.25);}}
are arbitrary tensors at $x=\pm \infty$. Due to the extensivity of energy $H_l(x_i)$ and 
$H_r(x_i)$ both contain infinities which, for numerical computations, have to be regularized 
\cite{haegeman_time-dependent_2011}, as will be detailed in the following. 
To ease up notation we use the abbrevation
\begin{align}
  &\mathcal{T}_{l/r}(x_i)\equiv\mathcal{P}e^{\int_{x_i}^{x_i+L}T_{l/r}(x)dx}\nonumber\\
  &T_{l/r}(x)=Q_{l/r}(x)\otimes\id+\id\otimes Q^*_{l/r}(x) + R_{l/r}(x)\otimes R^*_{l/r}(x)\nonumber
\end{align}
in the following.
For a periodic system, the calculation of $\bra{H_l(x_i)}$ and $\ket{H_r(x_i)}$ can be simplified to
\begin{align}
  \bra{H_l(x_i)}=\bra{h_l(x_i)}\sum_{n\in \mathbbm{N}}[\mathcal{T}_l(x_i)]^n\label{eq:geomseries1}\\
  \ket{H_r(x_i)}=\sum_{n\in \mathbbm{N}}[\mathcal{T}_r(x_i)]^n\ket{h_r(x_i)}\label{eq:geomseries2}.
\end{align}
The hermitian matrices $\bra{h_l(x_i)}$ and $\ket{h_r(x_i)}$ 
are defined as the contraction of an infinite tensor network of length $L$ of the form
\begin{equation}
    \begin{tikzpicture}[baseline={([yshift=0pt]X.base)},every node/.style={scale=1.0},scale=1.0]
      \draw (-1.6,0.0) node (X) {$\bra{h_l(x_i)}=$};
    \end{tikzpicture}
    \begin{tikzpicture}[baseline={([yshift=0pt]X.base)},every node/.style={scale=0.55},scale=1.35]
      % bottom row of rectangles
      \draw (1.6,0.4) node (X) {$A_{l}({{x_i-}L+}\eps)$};
      \draw (0.4,0.4) node (X) {$A_{l}({x_i-}L)$};
      \draw (0.0,0.4) to [out=180,in=180] (0,2.4);
      \draw[rounded corners] (0,0) rectangle (0.8,0.8);
      \draw[rounded corners] (1.2,0) rectangle (2.0,0.8);
      
      % top row of rectangles
      \draw (0.4,2.4) node (X) {$A_{l}({x_i-}L)$};
      \draw (1.6,2.4) node (X) {$A_{l}({{x_i-}L+}\eps)$};
      
      \draw[rounded corners] (0,2.0) rectangle (0.8,2.8);
      \draw[rounded corners] (1.2,2.0) rectangle (2.0,2.8);
      
      % center row of rectangles
      \draw[fill=white,rounded corners] (0,1.0) rectangle (0.8,1.8);
      \draw[fill=white,rounded corners] (1.2,1.0) rectangle (2.0,1.8);
      
      \draw (0.4,1.4) node (X) {${x_{i}-}L$};
      \draw (1.6,1.4) node (X) {${{x_{i}-L}+}\eps$};

      % center horizontal lines 
      \draw (0.8,1.4) -- (1.2,1.4);
      \draw (2.0,1.4) -- (2.4,1.4);
      
      % top horizontal lines 
      \draw (0.8,2.4) -- (1.2,2.4);
      \draw (2.0,2.4) -- (2.4,2.4);
      
      % bottom horizontal lines 
      \draw (0.8,0.4) -- (1.2,0.4);
      \draw (2.0,0.4) -- (2.4,0.4);
      
      % bottom vertical lines 
      \draw (0.4,0.8) -- (0.4,1.0);
      \draw (1.6,0.8) -- (1.6,1.0);
      
      % top vertical lines 
      \draw (0.4,1.8) -- (0.4,2.0);
      \draw (1.6,1.8) -- (1.6,2.0);
    \end{tikzpicture}
    \begin{tikzpicture}[baseline={([yshift=0pt]X.base)},every node/.style={scale=1.0},scale=1.0]
      \draw (-1.6,0.0) node (X) {$\dots$};
    \end{tikzpicture}
    \begin{tikzpicture}[baseline={([yshift=0pt]X.base)},every node/.style={scale=0.55},scale=1.35]
      % bottom row of rectangles
      \draw (1.6,0.4) node (X) {$A_{l}({x_i})$};
      \draw (2.0,0.4) -- (2.2,0.4);
      \draw (2.0,2.4) -- (2.2,2.4);
      \draw[rounded corners] (1.2,0) rectangle (2.0,0.8);
      
      % top row of rectangles
      \draw (1.6,2.4) node (X) {$A_{l}({x_i})$};
      
      \draw[rounded corners] (1.2,2.0) rectangle (2.0,2.8);
      
      % center row of rectangles
      \draw[fill=white,rounded corners] (1.2,1.0) rectangle (2.0,1.8);
      
      \draw (1.6,1.4) node (X) {${x_{i}+}L$};

      % center horizontal lines 
      \draw (0.8,1.4) -- (1.2,1.4);
      
      % top horizontal lines 
      \draw (0.8,2.4) -- (1.2,2.4);
      
      % bottom horizontal lines 
      \draw (0.8,0.4) -- (1.2,0.4);
      
      % bottom vertical lines 
      \draw (1.6,0.8) -- (1.6,1.0);
      
      % top vertical lines 
      \draw (1.6,1.8) -- (1.6,2.0);

    \end{tikzpicture}
    \nonumber
\end{equation}

\begin{equation}
  \ket{h_r(x_i)}=
    \begin{tikzpicture}[baseline = (X.base),every node/.style={scale=0.60},scale=1.2]
      % bottom row of rectangles
      \draw (1.6,0.4) node (X) {$A_{r}({x_i+}\eps)$};
      \draw (0.4,0.4) node (X) {$A_{r}({x_i}$};
      \draw (0,0.4) -- (-0.3,0.4);
      \draw (0,2.4) -- (-0.3,2.4);
      \draw[rounded corners] (0,0) rectangle (0.8,0.8);
      \draw[rounded corners] (1.2,0) rectangle (2.0,0.8);
      
      % top row of rectangles
      \draw (0.4,2.4) node (X) {$A_{r}({x_i})$};
      \draw (1.6,2.4) node (X) {$A_{r}({x_i+}\eps)$};
      
      \draw[rounded corners] (0,2.0) rectangle (0.8,2.8);
      \draw[rounded corners] (1.2,2.0) rectangle (2.0,2.8);
      
      % center row of rectangles
      \draw[fill=white,rounded corners] (0,1.0) rectangle (0.8,1.8);
      \draw[fill=white,rounded corners] (1.2,1.0) rectangle (2.0,1.8);
      
      \draw (0.4,1.4) node (X) {${x_{i}}$};
      \draw (1.6,1.4) node (X) {${x_{i}+}\eps$};

      % center horizontal lines 
      \draw (0.8,1.4) -- (1.2,1.4);
      \draw (2.0,1.4) -- (2.4,1.4);
      
      % top horizontal lines 
      \draw (0.8,2.4) -- (1.2,2.4);
      \draw (2.0,2.4) -- (2.4,2.4);
      
      % bottom horizontal lines 
      \draw (0.8,0.4) -- (1.2,0.4);
      \draw (2.0,0.4) -- (2.4,0.4);
      
      % bottom vertical lines 
      \draw (0.4,0.8) -- (0.4,1.0);
      \draw (1.6,0.8) -- (1.6,1.0);
      
      % top vertical lines 
      \draw (0.4,1.8) -- (0.4,2.0);
      \draw (1.6,1.8) -- (1.6,2.0);
    \end{tikzpicture}
    \dots
    \begin{tikzpicture}[baseline = (X.base),every node/.style={scale=0.60},scale=1.2]
      % bottom row of rectangles
      \draw (1.6,0.4) node (X) {$A_{r}({x_i+}L)$};
      \draw (2.0,0.4) to [out=0,in=0] (2.0,2.4);
      \draw[rounded corners] (1.2,0) rectangle (2.0,0.8);
      
      % top row of rectangles
      \draw (1.6,2.4) node (X) {$A_{r}({x_i+}L)$};
      
      \draw[rounded corners] (1.2,2.0) rectangle (2.0,2.8);
      
      % center row of rectangles
      \draw[fill=white,rounded corners] (1.2,1.0) rectangle (2.0,1.8);
      
      \draw (1.6,1.4) node (X) {${x_{i}+}L$};

      % center horizontal lines 
      \draw (0.8,1.4) -- (1.2,1.4);
      
      % top horizontal lines 
      \draw (0.8,2.4) -- (1.2,2.4);
      
      % bottom horizontal lines 
      \draw (0.8,0.4) -- (1.2,0.4);
      
      % bottom vertical lines 
      \draw (1.6,0.8) -- (1.6,1.0);
      
      % top vertical lines 
      \draw (1.6,1.8) -- (1.6,2.0);
    \end{tikzpicture}
    \nonumber.
\end{equation}
They are called 
the effective unit-cell Hamiltonians, i.e. the Hamiltonian operator
of a unit-cell projected into an effective, orthonormal basis set of the left respectively right infinite half-chain
(given by left and right isometric tensors $A^{\sigma}_{l/r}(x)$). The contraction of these networks is equivalent to 
solving a boundary-value problem for a system of ordinary differential equations for the matrices 
$\bra{h_l(x)}$ and $\ket{h_r(x)}$,
in analogy to the evolution of e.g. $\bra{l(x)}$ with the transfer operator $\mathcal{T}(x,y)$.
In semi-graphical notation, these differential equations are given by

\begin{widetext}
\begin{equation}\label{eq:hlode}
  \begin{tikzpicture}[baseline = (X.base),every node/.style={scale=1.0},scale=1.0]
    \draw (-0.35,0.2) to [out=90,in=180] (-0.15,0.5);
    \draw (-0.35,-0.2) to [out=270,in=180] (-0.15,-0.5);
    \draw[rounded corners] (-0.75,-0.2) rectangle (0.1,0.2);
    \draw (-0.5,0.0) node (X) {
      $\frac{d}{dx}h_l(x)$
    };
    \draw (0.35,0.0) node (X) {=};
  \end{tikzpicture} 
  \begin{tikzpicture}[baseline={([yshift=12pt]X.base)},every node/.style={scale=1.0},scale=1.0,xscale=1.7]
    \draw (-0.2,0.0) node (X) {$\frac{1}{2m}$};
    \draw[rounded corners] (0.25,0.1) rectangle (2.5,0.65);
    \draw (2.5,0.375) -- (2.65,0.375);
    \draw (1.35,0.4) node (X) {
      $[Q_l(x),R_l(x)]+\frac{dR_l(x)}{dx}$
    };
    \draw[rounded corners] (0.25,-0.15) rectangle (2.55,-0.7);
    \draw (0.25,0.375) to [out=180,in=180] (0.25,-0.42);
    \draw (2.55,-0.425) -- (2.7,-0.42);
    \draw (1.4,-0.4) node (X) {
      $[Q_l(x)^*,R_l(x)^*]+\frac{dR_l(x)^*}{dx}$
    };
  \end{tikzpicture} 
  +
  \begin{tikzpicture}[baseline={([yshift=11pt]X.base)},every node/.style={scale=1.0},scale=1.0,xscale=2.0]
    \draw (0.85,0.0) node (X) {$g$};
    \draw[rounded corners] (1.15,0.1) rectangle (1.7,0.65);
    \draw (1.7,0.375)--(1.85,0.375);
    \draw (1.4,0.4) node (X) {
      $R_l^2(x)$
    };
    \draw[rounded corners] (1.15,-0.1) rectangle (1.7,-0.65);
    \draw (1.15,0.375) to [out=180,in=180] (1.15,-0.375);
    \draw (1.7,-0.375)--(1.85,-0.375);
    \draw (1.4,-0.4) node (X) {
      $R_l^{*2}(x)$
    };
  \end{tikzpicture} 
  +
  \begin{tikzpicture}[baseline={([yshift=11pt]X.base)},every node/.style={scale=1.0},scale=1.0,xscale=2.0]
    \draw (0.7,0.0) node (X) {$\mu(x)$};
    \draw[rounded corners] (1.15,0.1) rectangle (1.7,0.65);
    \draw (1.7,0.375)--(1.85,0.375);
    \draw (1.4,0.4) node (X) {
      $R_l(x)$
    };
    \draw[rounded corners] (1.15,-0.1) rectangle (1.7,-0.65);
    \draw (1.15,0.375) to [out=180,in=180] (1.15,-0.375);
    \draw (1.7,-0.375)--(1.85,-0.375);
    \draw (1.4,-0.4) node (X) {
      $R_l^*(x)$
    };
  \end{tikzpicture} 
  +
  \begin{tikzpicture}[baseline={([yshift=0pt]X.base)},every node/.style={scale=1.0},scale=1.0,xscale=1.0]
    \draw (1.1,0.2) to [out=90,in=180] (1.75,0.5);
    \draw (1.1,-0.2) to [out=270,in=180] (1.75,-0.5);
    \draw[rounded corners] (0.7,-0.2) rectangle (1.55,0.2);
    \draw (1.125,0.0) node (X) {
      $h_l(x)$
    };
    \draw[rounded corners] (1.75,-0.7) rectangle (2.7,0.7);
    \draw (2.2,0.0) node (X) {
      $T_l(x)$
    };
    \draw (2.7,0.5) -- (2.9,0.5);    
    \draw (2.7,-0.5) -- (2.9,-0.5);
  \end{tikzpicture} 
\end{equation}

\begin{equation}\label{eq:hrode}
  \begin{tikzpicture}[baseline = (X.base),every node/.style={scale=1.0},scale=1.0]
    \draw (-0.35,0.2) to [out=90,in=0] (-0.55,0.5);
    \draw (-0.35,-0.2) to [out=270,in=0] (-0.55,-0.5);
    \draw[rounded corners] (-0.75,-0.2) rectangle (0.1,0.2);
    \draw (-0.5,0.0) node (X) {
      $\frac{d}{dx}h_r(x)$
    };
    \draw (0.35,0.0) node (X) {=};
  \end{tikzpicture} 
  \begin{tikzpicture}[baseline={([yshift=12pt]X.base)},every node/.style={scale=1.0},scale=1.0,xscale=1.7]
    \draw (-0.2,0.0) node (X) {$\frac{1}{2m}$};
    \draw[rounded corners] (0.25,0.1) rectangle (2.6,0.65);
    \draw (0.1,0.375) -- (0.25,0.375);    
    \draw (1.36,0.4) node (X) {
      $[Q_r(x),R_r(x)]+\frac{dR_r(x)}{dx}$
    };
    \draw[rounded corners] (0.25,-0.15) rectangle (2.6,-0.7);
    \draw (0.1,-0.425) -- (0.25,-0.42);    

    \draw (2.6,0.375) to [out=0,in=0] (2.6,-0.42);
    \draw (1.45,-0.4) node (X) {
      $[Q_r(x)^*,R_r(x)^*]+\frac{dR_r(x)^*}{dx}$
    };
  \end{tikzpicture} 
  +
  \begin{tikzpicture}[baseline={([yshift=11pt]X.base)},every node/.style={scale=1.0},scale=1.0,xscale=2.0]
    \draw (0.85,0.0) node (X) {$g$};
    \draw[rounded corners] (1.15,0.1) rectangle (1.7,0.65);
    \draw (1.0,0.375)--(1.15,0.375);    
    \draw (1.4,0.4) node (X) {
      $R_r^2(x)$
    };
    \draw[rounded corners] (1.15,-0.1) rectangle (1.7,-0.65);
    \draw (1.0,-0.375)--(1.15,-0.375);    
    \draw (1.7,0.375) to [out=0,in=0] (1.7,-0.375);
    \draw (1.4,-0.4) node (X) {
      $R_r^{*2}(x)$
    };
  \end{tikzpicture} 
  +
  \begin{tikzpicture}[baseline={([yshift=11pt]X.base)},every node/.style={scale=1.0},scale=1.0,xscale=2.0]
    \draw (0.7,0.0) node (X) {$\mu(x)$};
    \draw[rounded corners] (1.15,0.1) rectangle (1.7,0.65);
    \draw (1.0,0.375)--(1.15,0.375);    
    \draw (1.4,0.4) node (X) {
      $R_r(x)$
    };
    \draw[rounded corners] (1.15,-0.1) rectangle (1.7,-0.65);
    \draw (1.0,-0.375)--(1.15,-0.375);
    \draw (1.7,0.375) to [out=0,in=0] (1.7,-0.375);
    \draw (1.4,-0.4) node (X) {
      $R_r^*(x)$
    };
  \end{tikzpicture} 
  +
  \begin{tikzpicture}[baseline={([yshift=0pt]X.base)},every node/.style={scale=1.0},scale=1.0,xscale=1.0]
    \draw (3.3,0.2) to [out=90,in=0] (2.7,0.5);
    \draw (3.3,-0.2) to [out=270,in=0] (2.7,-0.5);
    \draw[rounded corners] (2.9,-0.2) rectangle (3.85,0.2);
    \draw (3.4,0.0) node (X) {
      $h_r(x)$
    };
    \draw[rounded corners] (1.75,-0.7) rectangle (2.7,0.7);
    \draw (2.2,0.0) node (X) {
      $T_r(x)$
    };
    \draw (1.55,0.5) -- (1.75,0.5);    
    \draw (1.55,-0.5) -- (1.75,-0.5);
  \end{tikzpicture} 
\end{equation}
\end{widetext}
with $T_r\ket{h_r(x)}$ and $\bra{h_l(x)}T_l$ given by \Eq{eq:transopoder} and \Eq{eq:transopodel},
and boundary conditions
\begin{align}
  \bra{h_l(x_i-L)}=0\nonumber\\
  \ket{h_r(x_i+L)}=0\nonumber.
\end{align}
$\bra{h_l(x_i)}$ ($\ket{h_r(x_i)}$) is then obtained by evolving $\bra{h_l(x_i-L)}$ ($\ket{h_r(x_i+L)}$)
from ${x=x_i-}L$ ($x={x_i+}L$) to ${x=}x_i$.

After proper normalization of the spMPS, the unit-cell transfer 
operators $\mathcal{T}_l(x_i)$ and $\mathcal{T}_r(x_i)$ have the eigenmatrix $\bra{\mathbbm{1}}$ and
$\ket{\mathbbm{1}}$ to eigenvalue $\eta=1$. The geometric series \Eq{eq:geomseries1} 
and \Eq{eq:geomseries2} thus diverge and the sums cannot be 
performed trivially. The infinities in \Eq{eq:geomseries1} and \Eq{eq:geomseries2} are 
due to an infinite energy expectation value of
the cMPS $\ket{\Psi}$. Since this is just an infinite energy offset, 
one can savely subtract it from the Hamiltonian. One way of achieving this is by 
a proper regularization of the unit-cell transfer operator $\mathcal{T}_{l/r}(x_i)$ 
\cite{zauner-stauber_variational_2017}. 
This is done by projecting it 
into the subspace orthogonal to $\ket{\mathbbm{1}}\bra{l(x_i)}$ 
and $\ket{r(x_i)}\bra{\mathbbm{1}}$,
respectively, i.e. by replacing $\mathcal{T}_{l/r}(x_i)$ with 
\begin{align}
    \mathcal{T}_{l\perp}(x_i)&=\mathcal{T}_{l}(x_i)-\ket{r(x_i)}\bra{\mathbbm{1}}\nonumber\\
    \mathcal{T}_{r\perp}(x_i)&=\mathcal{T}_{r}(x_i)-\ket{\mathbbm{1}}\bra{l(x_i)}\nonumber\\
    \bra{h_l(x_i)}_{\perp}&=\bra{h_l(x_i)}-\braket{h_l(x_i)|r(x_i)}\bra{\mathbbm{1}}\nonumber\\
    \ket{h_r(x_i)}_{\perp}&=\ket{h_r(x_i)}-\braket{l(x_i)|h_r(x_i)}\ket{\mathbbm{1}}\nonumber.
\end{align}
We then solve for $\bra{H_l(x_i)}$ and $\ket{H_r(x_i)}$ by inverting the equations
\begin{align}
    \bra{h_l(x_i)}_{\perp}\frac{1}{\mathbbm{1}-\mathcal{T}_{l\perp}(x_i)}=\bra{H_l(x_i)}\label{eq:lgmres1}\\
    \frac{1}{\mathbbm{1}-\mathcal{T}_{r\perp}(x_i)}\ket{h_r(x_i)}_{\perp}=\ket{H_r(x_i)}\label{eq:lgmres2}
\end{align}
using a sparse solver, like e.g. the {\it lgmres} routine provided by the 
{\it scientific python} package.
In practice it is not necessary to solve \Eq{eq:lgmres1} and \Eq{eq:lgmres2} for each $x_i$. 
Instead, we first solve it for a single point, e.g. $x_i=0$, and then evolve $\bra{H_l(0)}$ 
and $\ket{H_r(0)}$ to all other points $x_i$ using \Eq{eq:hlode} and \Eq{eq:hrode}.

\end{document}